\documentclass[preprint, prd, nofootinbib, aps, showpacs, showkeys, preprintnumbers]{revtex4-1}
\usepackage{amsmath,amsfonts,amssymb,amsthm,amstext,amscd,eucal}
\usepackage[all]{xy}
\usepackage{epsfig}
\usepackage{amsmath}
\usepackage{amssymb}
\usepackage{graphicx}
\makeatletter \@addtoreset{equation}{section}

\newcommand{\ie}{{\em i.e. }}
\newcommand{\eg}{{\em e.g. }}

\newcommand{\be}{\begin{equation}}
\newcommand{\ee}{\end{equation}}
\newcommand{\bea}{\begin{eqnarray}}
\newcommand{\eea}{\end{eqnarray}}

\newcommand{\bse}{\begin{subequations}}
\newcommand{\ese}{\end{subequations}}
\newcommand{\bi}{\begin{itemize}}
\newcommand{\ei}{\end{itemize}}

\newcommand{\mpl}{M_{\rm pl}}

\def\tM{\tilde{\mathcal{M}}}

\def\Tr{  \mbox{Tr}   }

\newcommand{\mQ}{\mathcal{Q}}

\newcommand{\dre}{\delta \rho}
\newcommand{\dpe}{\delta P}
\newcommand{\dqe}{\delta q}
\def\YM2{\frac{g^2\phi^4}{a^4}}
\newcommand{\mM}{\mathcal{M}}
\newcommand{\mR}{\mathcal{R}}
\newcommand{\mH}{\mathcal{H}}

\def\kk{\frac{\kappa g^2\phi^4}{a^4}}
\def\kk{\frac{\kappa g^2\phi^4}{a^4}}
\def\dk{\frac{\kappa \dot\phi^2}{a^2}}
\def\dre_g{\delta\rho_g}
\def\dpe_g{\delta P_g}
\def\dqe_g{\delta q_g}
\def\dre{\delta\rho}
\def\dpe{\delta P}
\def\dqe{\delta q}
\def\mH{\mathcal{H}}

\def\YM1{\frac{\dot\phi^2}{a^2}}
\def\YM2{\frac{g^2\phi^4}{a^4}}
\begin{document}

\preprint{ IPM/P-2010/009}
\preprint{arXiv:1102.1932}
\vskip1cm
\title{Non-Abelian Gauge Field Inflation}

\author{A.~Maleknejad}
\email{azade@ipm.ir}\affiliation {Department of Physics,
Alzahra University
P. O. Box 19938, Tehran 91167, Iran, and\\
School of Physics, Institute for research in fundamental sciences
(IPM), P.O.Box 19395-5531, Tehran, Iran }

\author{M.~M.~Sheikh-Jabbari}
\email{jabbari@theory.ipm.ac.ir} \affiliation{School of Physics,
Institute for Research in Fundamental Sciences (IPM), P. O. Box
19395-5531, Tehran, Iran}

\date{\today}

\begin{abstract}
In \cite{gauge-flation} we introduced an inflationary scenario,
\emph{non-abelian gauge field inflation} or \emph{gauge-flation} for short, in which slow-roll
inflation is driven by non-Abelian gauge field minimally coupled to
gravity. We present a more detailed analysis, both numerical and analytical, of the gauge-flation.
By studying the phase diagrams of the theory,
we show that getting enough number of e-folds during a slow-roll
inflation is fairly robust to the choice of initial gauge field
values. In addition, we present a detailed analysis of the cosmic
perturbation theory in gauge-flation which has many special and interesting
features compared the standard scalar-driven inflationary models.
The specific gauge-flation model we study in this paper has two parameters,
a cutoff scale $\Lambda$ and the gauge coupling $g$. Fitting our results
with the current cosmological data fixes $\Lambda\sim 10 H \sim
10^{14}$ GeV ($H$ is the Hubble parameter) and $g\sim 10^{-3}$,
which are in the natural range of parameters in generic
particle physics beyond standard models. Our model also predicts a tensor-to-scalar ratio  $r>0.02$, in the range detectable by the Planck satellite.

\end{abstract}
\maketitle


\section{Introduction}

The idea of inflationary cosmology was originally proposed to provide a possible resolution to some of the theoretical problems of the bing-bang model for the early Universe cosmology \cite{Inflation-Books}. However, with the advancement of the cosmological observations and most notably the cosmic microwave background (CMB) observations \cite{Inflation-Books,Bassett-review},  the inflationary paradigm has received observational support and inflation is now considered an integral part of the standard model of cosmology with the following general picture. A patch of the early Universe which is a few Planck lengths in size under the gravitational effects of the matter present there undergoes a rapid (usually exponential) expansion, the inflationary period. The inflation ends while most of the energy content of the Universe is still concentrated in the field(s) driving  inflation, the inflaton field(s). This energy should now be transferred to the other fields and particles, the (beyond) standard model particles,  through the (p)reheating process.
The rest of the picture is that of the standard hot big-bang scenario, with radiation dominated, matter dominated and finally the dark energy dominated era that we live in.

In the absence of a direct observation for the primordial gravity waves, one of the main standing issues in inflation is what is the Hubble parameter during inflation $H$, or the energy density of the inflaton field(s). With the current observations, and within the slow-roll inflation scenario, the preferred scale is $H\lesssim 10^{-5}\mpl$, where $\mpl\equiv (8\pi G_N)^{-1/2}= 2.43 \times 10^{18}$ GeV is the reduced Planck mass. On the other hand, according to the lore in beyond standard particle physics models, the supersymmetric grand unified theories (SUSY GUTs) \cite{Dine-Book}, the unification scale is around $10^{16}$ GeV, suggesting that inflationary model building should be sought for within
various corners of such models. If so, the SUSY GUT setting will also provide a natural arena for building the (p)reheating models.

Almost all of inflationary models or model building ideas that appear in the literature use one or more scalar fields with a suitable potential to provide for the matter field inducing the inflationary expansion of the early Universe.
The choice of scalar fields is made primarily because we work within the homogeneous and isotropic Friedmann-Robertson-Walker (FRW) cosmology and that turning on spinor or gauge fields generically violates these symmetries. Another reason is that, from the model building viewpoint, turning on potential for the scalar fields is easier than for other fields, whose interactions are generically fixed by gauge symmetries or renormalizability conditions. Building inflationary models within the SUSY GUTs  then amounts to exploring various corners of the theory/model  in search of flat enough potential which supports successful slow-roll inflation, the flatness of which is respected by the loop and quantum corrections. Such models  usually come under the D-term or F-term inflationary models \cite{Linde-Kallosh}.

Regardless of the details, non-Abelian gauge field theories are the widely accepted framework for building  particle physics models, and, in particular, beyond standard models and GUTs. In view of the ubiquitous appearance of non-Abelian gauge fields, one may explore the idea of using gauge fields as inflaton fields, the fields which get nonzero background value during inflation and drive the inflationary dynamics. One of the main obstacles in this regard is the vector nature of the gauge fields and that turning them on in the background will spoil the rotation symmetry.

A related scenario in which this problem was pointed out and addressed is ``vector inflation \cite{vector-inflation}.'' The idea in vector inflation, unlike ours, is to use vector fields and not gauge fields, as inflaton. In \cite{vector-inflation}, two possible ways were proposed to overcome the broken rotational invariance caused by the vector inflaton fields:
(1) introduce a large number of vectors each assuming a random direction in the 3D space, such that on the average we recover a rotational invariant background; or, (2) introduce three orthogonal vector fields of the same value which act as the triad of the spatial part of the spacetime, the ``triad method'' \cite{vector-inflation-3,Bento:1992wy}.
The other important obstacle in the way of driving inflation by vector fields is the exponential, $1/a(t)$ suppression of the massless vector fields in an inflationary background, causing inflation to end too fast. This problem has been overcome by adding nonminimal coupling to the gravity, usually a conformal mass type term \cite{vector-inflation,vector-inflation-2}. To have a successful inflation, however, this is not enough and one should add quite nontrivial potentials for the vector field \cite{vector-inflation,vector-inflation-2,{vector-inflation-3}}. Dealing with vector fields, and not gauge fields, may bring instabilities in the theory: the longitudinal mode of the vector field which has a ghost type kinetic term and is not dynamical at tree level, in the absence of gauge invariance, can and will, become dynamical once quantum (loop) effects are taken into account. This latter will cause ghost instability, if we were studying the theory on a flat background. It has been argued that such instabilities can persist in the inflationary background too \cite{vector-inflation-loophole}; see, however, \cite{Lyth-vector} for a counter argument. In any case the instability issue of vector inflationary models seems not to be settled yet.

In order not to face the above issue one should build a ``gauge invariant vector inflation.'' One can easily observe that it is not possible to get a successful inflation with some number of $U(1)$ gauge fields. The other option is to consider non-Abelian gauge theories. The ``triad method'' mentioned above is naturally realized within the non-Abelian gauge symmetry setting, irrespective of the gauge group in question. We then face the second obstacle, the $1/a(t)$ suppression. This may be achieved by changing the gravity theory, considering Yang-Mills action coupled to $F(R)$ modified gravity \cite{YM-F(R)}, or considering Einstein gravity coupled to a generic (not necessarily Yang-Mills) gauge theory action. This latter is the idea we will explore in this work. We should stress that, as will become clear, our approach and that of \cite{YM-F(R)} are basically different. Using non-Abelian gauge fields has another advantage that, due to the presence of $[A_\mu,A_\nu]$ term in the gauge field strength $F_{\mu\nu}$, it naturally leads to a ``potential'' term for the gauge fields which, upon a suitable choice of the gauge theory action,  can be used to overcome the $1/a(t)$ suppression problem mentioned above.

In this work, we present a detailed discussion and analysis  of gauge-flation, inflation driven by non-Abelian gauge fields, which we introduced in \cite{gauge-flation}. In section \ref{setup-section}, we show how the rotation symmetry breaking can be compensated by the $SU(2)$ (sub)group of the global part of non-Abelian gauge symmetry and how one can introduce a combination of the gauge field components which effectively behaves as spacetime scalar field (on the FRW background); and that there is a consistent truncation from the classical phase space of the non-Abelian gauge theory to the sector which only involves this scalar field.

Setting the stage, in section \ref{the-model-analytic}, we choose a specific action for the gauge theory that is Yang-Mills plus a specific $F^4$ term which can come from specific (one) loop corrections to the gauge theory. In this work, however, we adopt a phenomenological viewpoint and
choose this specific $F^4$ term primarily for the purpose of inflationary model building. The important point of providing field theoretical justifications for this $F^4$ term will be briefly discussed in the discussion section and dealt with in more detail in an upcoming publication. Our model has hence two parameters, the gauge coupling $g$ and the ceofficient of this specific $F^4$ term $\kappa$. These two parameters will be determined only by focusing on the considerations coming from cosmological observations. In this section we present an analytic study of the inflationary dynamics of our gauge-flation model and show that the model allows for a successful slow-roll inflationary period which leads to enough number of e-folds. In section \ref{numerical-analysis-section}, we present the diagrams and graphs for the numerical analysis of the gauge-flation model. Our numerical analysis reveals that the  classical slow-roll inflationary trajectory is fairly robust to the choice of initial conditions.

Having studied the classical inflationary dynamics, in section \ref{GF-cosmic-pert-theory-section}, we turn to the question of cosmic perturbation theory in the gauge-flation. Because of the existence of other components of the gauge fields which has been turned off in the classical inflationary background, the situation here is considerably different than the standard cosmic perturbation theory developed in the literature. We hence first develop the cosmic perturbation theory for our model, discuss its subtleties and novelties; we discuss the scalar, vector and tensor perturbations, their power spectra and the spectral tilts. In section \ref{testing-the-model}, after completing the analysis of the model, we confront our model with the available cosmological and CMB data. We show that indeed it is possible to get a successful inflationary model with the gauge-flation setup. In the last section we summarize our results and make concluding remarks. In three appendices we have gathered some technical details of the cosmic perturbation theory.

\textbf{Note added:} After we published this work the paper \cite{AMW} appeared which prompted us to recheck the cosmic perturbation theory analysis of our earlier version, especially in the tensor mode perturbations. We have now corrected and improved our cosmic perturbation theory analysis. We have also improved our analysis in comparing the results of our gauge-flation model with the CMB and other cosmic data. These analysis have also appeared in  the review article \cite{gauge-field-review}.

\section{The setup}\label{setup-section}

In this section we first demonstrate how the rotation symmetry is retained in the gauge-flation and then discuss truncation to the scalar sector. Here we will consider an $SU(2)$ gauge theory with gauge fields $A_\mu^a$ where $a=1,2,3$ label the gauge algebra indices and $\mu=0,1,2,3$ the spacetime indices, the temporal components will be denoted by $A_0^a$ and the spatial components by $A_{i}^a$. Although we focus on the $SU(2)$ gauge theory, our analysis holds for any non-Abelian gauge group $G$, as any non-Abelian  group  always has an $SU(2)$ subgroup.

We will consider gauge- and Lorentz-invariant theories, where the gravity part is the usual Einstein-Hilbert action and the Lagrangian of the gauge theory part, which is \emph{minimally coupled} to gravity, is of the form
${\cal L}=\mathcal{L} (F^a_{\mu\nu}; g_{\mu\nu})$, where $F^a_{\mu\nu}$ is the gauge field strength
\be\label{F-definition}%
F^a_{~\mu\nu}=\partial_\mu A^a_{~\nu}-\partial_\nu A^a_{~\mu}-g\epsilon^a_{~bc}A^b_{~\mu}A^c_{~\nu}\ .
\ee
(For a generic gauge group, $\epsilon^{abc}$ should be replaced with the structure constant of that group.)
Under the action of gauge transformation $U=exp(-\lambda_aT^a)$, where $T^a$ are generators of the $su(2)$ algebra,
\be%
[T^a,T^b]=i\epsilon^{ab}_{\ \ c}T^c\,,
\ee%
$A^a_{~\mu}$ transforms as
\be \label{ug}
A_{\mu}\rightarrow UA_{\mu}U^\dag-\frac{1}{g}U\partial_\mu U^\dag\,.
\ee
Therefore, out of 12 components of  $A^a_{~\mu}$, nine are physical and three are gauge freedoms, which may be removed by a suitable choice of gauge parameter $\lambda_a$. Since we are interested in isotropic and homogeneous  FRW cosmology, the temporal gauge
\be
A^a_{~0}=0\,,
\ee
appears to be a suitable gauge fixing. This fixes the gauge symmetry \eqref{ug}, up to the \emph{global, time independent} $SU(2)$ gauge transformations.
This global $SU(2)$ is the key to  restoring the rotation symmetry in the presence of the background gauge fields. We identify this $SU(2)$ with the three-dimensional rotations of the FRW background and, since the physical observables of the gauge fields are defined up to gauge transformations (or in other words, only gauge-invariant combinations are physical observables) the rotation symmetry may be preserved. This latter is done by turning on a specific gauge field configuration in which this identification can be made.

In order to see the above in a more technical language, consider the background FRW metric
\be \label{FRW}
ds^2=-dt^2+a^2(t)\delta_{ij}dx^idx^j\,,
\ee
where $i,j=1,2,3$ denote the indices along the spacelike three-dimensional hypersurface $\Sigma$, whose metric
is chosen to be $a^2\delta_{ij}$. By choosing  the (comoving cosmic) time direction,  metric on $\Sigma$ is then defined up to 3D foliation preserving diffeomorphisms. If we denote the metric on constant time hypersurfaces $\Sigma$ by $q_{ij}$, we can introduce a set of three vector fields, $\{e^a_{~i}(p)\}$, the triads, spanning the local Euclidian tangent space $T\Sigma_p$ to the $\Sigma$ at the given point $p$.
The triads satisfy the following orthonormality relations
\be\label{triad-ortho}%
q_{ij}=e^a_{~i}e^b_{~j}\delta_{ab}\,,\qquad
\delta^{ab}=e^a_{~i}e^b_{~j}q^{ij}\,.
\ee
The triads are then defined up to local 3D translations and rotations, which act on the ``local indices''$a,b$.
In particular the triads which are
related to  each other by  local rotations $\Lambda^a_{~b}\in SO(3)$
\be\label{rotation-triad}
e^a_{~i}\rightarrow \tilde{e}^a_{~i}=\Lambda^a_{~b}(p) e^b_{~i},
\ee
at each point $p$, lead to the same metric $q_{ij}$. The elements of this local $SO(3)$ may be
expressed in terms of $su(2)$ generators $T^a$ as $\Lambda=\exp(-\lambda_aT^a)$.
There are (infinitely) many possibilities for $e^a_{~i}$ for the FRW metric \eqref{FRW}, and one
obvious choice is
\be
e^a_{~i}=a(t)\delta^a_{i},
\ee
which identifies the space coordinate indices $i, j, k,\cdots$, with the local frame indices $a, b, c,\cdots$.

We may now readily identify the remaining global $SU(2)$ gauge symmetry with the global part of the 3D rotation symmetry \eqref{rotation-triad}. This can be done through the following ansatz%
\bea  \label{A-ansatz-background}
A^a_{~i}=\psi(t)e^a_{~i}=a(t)\psi(t)\delta^a_i , %
\eea%
where under both
of 3D diffeomorphisms and gauge transformations, $\psi(t)$ acts as a
genuine scaler field. Technically, the ansatz \eqref{A-ansatz-background} identifies the combination of the gauge fields for which the rotation symmetry violation caused by turning on vector (gauge) fields in the background is compensated for (or undone by) the gauge transformations, leaving us with rotationally invariant background.

As a result of this identification  the energy-momentum tensor produced by the gauge field configuration \eqref{A-ansatz-background} takes the form of a standard homogeneous, isotropic perfect fluid
\be
T^{\mu}_{\,\,\nu}=diag(-\rho,P,P,P)\,.
\ee
To see this, consider a general gauge and Lorentz-invariant gauge field Lagrangian density
${\cal L}=\mathcal{L}(F^a_{\mu\nu}; g_{\mu\nu})$. The corresponding energy-momentum tensor is given by
\be\label{Tmunu}
T_{\mu\nu}\equiv\frac{-2}{\sqrt{-\mathfrak{g}}}\frac{\delta(\sqrt{-\mathfrak{g}}\mathcal{L})}{\delta
g^{\mu\nu}}=2\frac{\delta\mathcal{L}}{\delta
F_{~\sigma}^{a~\mu}}F^a_{~\sigma\nu}+g_{\mu\nu}\mathcal{L}\,.
\ee
To compute $T_{\mu\nu}$, we need to first calculate the field strength $F_{\mu\nu}^a$ for $A_\mu^a$ in the temporal gauge $A^a_{~0}=0$, and for the field configuration of \eqref{A-ansatz-background}:
\be\label{F-background}%
\begin{split}
F^a_{~0i}&=\dot{\phi}\delta^a_i\,,\cr
F^a_{~ij}&=-g\phi^2\epsilon^a_{~ij}\,,
\end{split}
\ee%
where dot denotes derivative with respect to the comoving time $t$ and for the ease of notation we have introduced%
\be\label{phi-psi}%
\phi\equiv a(t)\psi(t)\,.
\ee
(Note that $\phi$, unlike $\psi$, is not a scalar.)
It is now straightforward to calculate energy density $\rho$ and pressure $P$, in terms of  $\phi$ and its time derivatives. Plugging \eqref{F-background} into \eqref{Tmunu} yields%
\bea\label{rho}
\rho&=&\frac{\partial\mathcal{L}_{red.}}{\partial\dot{\phi}}\dot{\phi}-\mathcal{L}_{red.},\\
\label{P}P &=& \frac{\partial(a^3 {\cal L}_{red.})}{\partial a^3},
\eea
where ${\cal L}_{red.}$ is the \emph{reduced Lagrangian} density, which is obtained from calculating
${\cal L}(F_{\mu\nu}^a; g_{\mu\nu})$ for field strengths $F_{\mu\nu}^a$ given in \eqref{F-background} and FRW metric
\eqref{FRW}.

One can check that ${\cal L}_{red.}$ is the true reduced Lagrangian   for the reduced phase space of the field configurations in the ansatz \eqref{A-ansatz-background} (and in the temporal gauge).
In order to do this, one can show that the gauge field equations of motion%
\be%
D_\mu \frac{\partial\mathcal{L}}{\partial F^a_{\mu\nu}}=0\,, %
\ee%
where $D_\mu$ is the gauge covariant derivative, \emph{(i)} allow for a solution
of the form \eqref{A-ansatz-background} and, \emph{(ii)} once evaluated on
the ansatz \eqref{A-ansatz-background} become equivalent to the
equation of motion obtained from the reduced Lagrangian
$\mathcal{L}_{red.}(\dot{\phi},\phi;a(t))$
\be\label{red-e.o.m}%
\frac{d}{a^3 dt}(a^3\frac{\partial {\cal L}_{red.}}{\partial
\dot\phi})-\frac{\partial {\cal L}_{red.}}{\partial \phi}=0 \,.
\ee%
In technical terms, there exists a consistent truncation of the gauge field theory to the  sector
specified by the scalar field $\psi$ (or $\phi$). In the next section we will study the cosmology of this reduced Lagrangian, with a specific choice for the gauge field theory action.

\section{A specific Gauge-flation model, analytic treatment}\label{the-model-analytic}

In the previous section we showed how homogeneity and isotropy can be preserved in a specific sector of any non-Abelian gauge field theory. In this section we couple the gauge theory to gravity and search for gauge field theories which can lead to a successful inflationary background. The first obvious choice is Yang-Mills action
minimally coupled to Einstein gravity. This will not lead to an
inflating system  because, as a result of scaling
invariance of Yang-Mills action, one immediately obtains $P=\rho/3$
and $\rho\geq 0$, and in order to have inflation we should have $\rho+3P<0$. So, we need to consider modifications to
Yang-Mills.

As will become
clear momentarily, one such appropriate choice involving $F^4$ terms
is
\be\label{The-model}%
S=\int
d^4x\sqrt{-{g}}\left[-\frac{R}{2}-\frac{1}{4}F^a_{~\mu\nu}F_a^{~\mu\nu}+\frac{\kappa}{384
}
(\epsilon^{\mu\nu\lambda\sigma}F^a_{~\mu\nu}F^a_{~\lambda\sigma})^2\right]\,
\ee
where we have set $8\pi G\equiv \mpl^{-2}=1$ and
$\epsilon^{\mu\nu\lambda\sigma}$ is the totally antisymmetric
tensor. We stress that this specific $F^4$ term is chosen only for inflationary model building purposes and, since the contribution
of this term to the energy-momentum tensor has the equation of
state $P=-\rho$, it is perfect for driving inflationary
dynamics. The justification of this term within a more rigorous quantum gauge field theory setting will be briefly discussed in section VII. (To respect the weak energy condition for the $F^4$ term,
we choose $\kappa$ to be positive.)

 The reduced (effective)
Lagrangian is obtained from evaluating \eqref{The-model} for the
ansatz \eqref{A-ansatz-background} \be
\mathcal{L}_{red}=\frac{3}{2}(\frac{\dot{\phi}^2}{a^2}-\frac{g^2\phi^4}{a^4}+\kappa
\frac{g^2\phi^4\dot{\phi}^2}{a^6})\,. \ee The energy density $\rho$
and pressure $P$ are \bea\label{rho-background}
\rho=\frac{3}{2}(\frac{\dot{\phi}^2}{a^2}+\frac{g^2\phi^4}{a^4}+\kappa
\frac{g^2\phi^4\dot{\phi}^2}{a^6})\, ,\\ \label{P-background}
P=\frac{1}{2}(\frac{\dot{\phi}^2}{a^2}+\frac{g^2\phi^4}{a^4}-3\kappa
\frac{g^2\phi^4\dot{\phi}^2}{a^6})\, . \eea As we see $\rho$ and $P$
have Yang-Mills parts and the $F^4$ parts, the $\kappa$ terms. If we
denote the Yang-Mills contribution to $\rho$ by $\rho_{_{YM}}$ and
the $F^4$ contribution by $\rho_{\kappa}$, \ie,
\be\label{rho0-rho1}%
\rho_{_{YM}}=\frac{3}{2}(\frac{\dot{\phi}^2}{a^2}+\frac{g^2\phi^4}{a^4})\,,\qquad
\rho_{\kappa}=\frac32\frac{\kappa g^2\phi^4\dot{\phi}^2}{a^6}\,,
\ee%
then%
\be%
\rho=\rho_{_{YM}}+\rho_{\kappa}\,,\qquad
P=\frac13\rho_{_{YM}}-\rho_\kappa\,.
\ee%
Field equations, the Friedmann equations and $\phi$ equation of motion,   are then obtained as%
\bea \label{cosm1}
H^2=\frac{1}{2}&(&\frac{\dot{\phi}^2}{a^2}+\frac{g^2\phi^4}{a^4}+\kappa \frac{g^2\phi^4\dot{\phi}^2}{a^6})
\,,\\ \label{cosm2}
\dot{H}~&=&-(\frac{\dot{\phi}^2}{a^2}+\frac{g^2\phi^4}{a^4})\,,\\
\label{phi-eom}
(1+\kappa\frac{g^2\phi^4}{a^4})\frac{\ddot{\phi}}{a}&+&(1+\kappa\frac{\dot{\phi}^2}{a^2})\frac{2g^2\phi^3}{a^3}+
(1-3\kappa\frac{g^2\phi^4}{a^4})\frac{H\dot{\phi}}{a}=0\,.
\eea

We start our analysis by exploring the possibility of slow-roll dynamics. To this end it is useful to introduce slow-roll parameters\footnote{We note that our definition of slow-roll parameters $\epsilon,\ \eta$ for the standard  single scalar inflationary theory $L=\frac12\dot\varphi^2-V(\varphi)$ reduces to \cite{Inflation-Books} $\epsilon=\frac{\mpl^2}{2}\left(\frac{V'}{V}\right)^2,\ \eta=\mpl^2\frac{V''}{V}$.}%
\be\label{slow-roll-parameters-def}
\epsilon\equiv -\frac{\dot{H}}{H^2}\,,\qquad \eta\equiv -\frac{\ddot{H}}{2\dot{H}H}\,,
\ee
where $\epsilon$ is the standard slow-roll parameter and $\eta$ is related to the time derivative of $\epsilon$ as%
\be \label{epsieta}
\eta=\epsilon-\frac{\dot{\epsilon}}{2H\epsilon}\,.
\ee
Therefore, to have a sensible slow-roll dynamics one should demand $\epsilon,\ \eta\ll 1$. Using the Friedmann equations \eqref{cosm1} and \eqref{cosm2} and definitions \eqref{rho0-rho1} we have%
\be\label{epsilon-rho0-rho1}
\epsilon= \frac{2\rho_{_{YM}}}{\rho_{_{YM}}+\rho_\kappa}\,.
\ee%
That is, to have slow-roll the $\kappa$-term contribution
$\rho_\kappa$ should dominate over the Yang-Mills contributions
$\rho_{_{YM}}$, or $\rho_\kappa\gg \rho_{_{YM}}$. As we will see, the
time evolution will then increase $\rho_{_{YM}}$ with respect to
$\rho_\kappa$, and when $\rho_{_{YM}}\sim \rho_\kappa$, the
slow-roll inflation ends. Noting that
$\rho+3P=2(\rho_{_{YM}}-\rho_\kappa)$, inflation (accelerated
expansion phase) will end when $\rho_{_{YM}}>\rho_\kappa$.

For having slow-roll inflation, however, it is \emph{not} enough to make sure $\epsilon\ll 1$. For the latter, time-variations of $\epsilon$ and all the other physical dynamical variables of the problem, like $\eta$ and the $\psi$ field, must also remain small over a reasonably large period in time (to result in enough number of e-folds).
To measure this latter we define%
\be\label{delta-def}
\delta\equiv-\frac{\dot{\psi}}{H\psi}\,,
\ee%
in terms of which the equations \eqref{cosm1}-\eqref{phi-eom} take the form%
\bea\label{epsil}
\epsilon&=&2-\kappa g^2\psi^6(1-\delta)^2, \\ \label{tilde-eta}
\eta&=&\epsilon-(2-\epsilon)\left[\frac{\dot{\delta}}{H(1-\delta)\epsilon}+\frac{3\delta}{\epsilon}\right]\,.
\eea
Comparing \eqref{epsieta} and \eqref{tilde-eta} we learn that to have a successful slow-roll, $\dot\epsilon\sim H\epsilon^2$ and $\eta\sim\epsilon$, we should demand that $\delta\sim \epsilon^2$. Explicitly, the equations of motion \eqref{cosm1}, \eqref{cosm2} and \eqref{phi-eom} admit the solution \footnote{Our numerical analysis reveals that even if we start with $\dot\delta/(H\delta)\sim {\cal O}(1)$, while $\psi_i^2\sim \epsilon\ll 1$, after a short time it becomes very small and hence for almost all the inflationary period we may confidently use $\delta\simeq \frac{\gamma}{6(\gamma+1)}\epsilon^2$. See section \ref{numerical-analysis-section} for a more detailed discussion.
}%
\bea \label{epsilon-x}
\epsilon&\simeq&\psi^2(\gamma+1),\\
\label{eta-x}
\eta &\simeq&\psi^2\quad \Rightarrow \quad (3+\frac{\dot\delta}{H\delta})\delta\simeq\frac{\gamma}{2(\gamma+1)}\epsilon^2\,,\\
\label{kappa-x}
\kappa&\simeq&\frac{(2-\epsilon)(\gamma+1)^3}{g^2\epsilon^3}, %
\eea%
where $\simeq$ means equality to first order in slow-roll parameter $\epsilon$ and \footnote{Note that all the dimensionful parameters, \ie $H,\psi$ and $\kappa$, are measured in units of $\mpl$; $H,\ \psi$ have dimension of energy while $\kappa$ has dimension of one-over-energy density.}
\be\label{x-def}
\gamma= \frac{g^2\psi^2}{H^2}\,,\qquad \textrm{or
equivalently}\qquad
H^2\simeq\frac{g^2\psi^4}{\epsilon-\psi^2}=\frac{g^2\epsilon}{\gamma(\gamma+1)}\,.%
\ee%
In the above $\gamma$ is a positive parameter which is slowly varying during slow-roll inflation.

Recalling \eqref{delta-def} and that $\delta\sim \epsilon^2$, \eqref{x-def} implies that $\gamma H^2$ remains almost a constant during the slow-roll inflation and hence \cite{gauge-flation}%
\be\label{epsilon-H}%
\frac{\epsilon}{\epsilon_i}\simeq\frac{\gamma+1}{\gamma_i+1}\,,\qquad \frac{\gamma}{\gamma_i}\simeq\frac{H_i^2}{H^2}\,,
\ee%
where $\epsilon_i,\ \gamma_i$ and $H_i$ are the values of these parameters at
the beginning of inflation.  As discussed the (slow-roll) inflation ends when $\epsilon=1$, where%
\be \gamma_f\simeq \frac{\gamma_i+1}{\epsilon_i}\,,\qquad
\frac{H_f^2}{H_i^2}\simeq \frac{\gamma_i}{\gamma_i+1}\ \epsilon_i\,.
\ee Using the above and \eqref{slow-roll-parameters-def} one can
compute the number of e-folds $N_e$ \bea\label{Ne}
N_{e}=\int_{t_i}^{t_f} Hdt=-\int_{H_i}^{H_f} \frac{dH}{\epsilon
H}\simeq \frac{\gamma_i+1}{2\epsilon_i}
\ln\frac{\gamma_i+1}{\gamma_i}\,. \eea

\section{Numerical analysis}\label{numerical-analysis-section}

As pointed out, our gauge-flation model has two parameters, the gauge coupling $g$ and the coefficient of the $F^4$ term $\kappa$. The degrees of freedom in the scalar sector of the model consists of the scalar field $\psi$ and the scale factor $a(t)$ and hence our solutions are specified by four initial values for these parameters and their time derivatives. These were parameterized by $H_i,\ \epsilon_i$ and $\psi_i$ and $\delta_i$ (or $\dot\psi_i$). The Friedmann equations, however, provide some relations between these parameters; assuming slow-roll dynamics these relations are \eqref{epsilon-x}-\eqref{kappa-x}. As a result each inflationary trajectory may  be specified by
the values of four parameters, $(g, \kappa;\ \psi_i, \dot\psi_i)$.
In what follows we present the results of the numerical analysis of the equations of motion \eqref{cosm1}, \eqref{cosm2} and \eqref{phi-eom}, for three sets of values for $(\psi_i,\dot\psi_i;\ g, \kappa)$.

\begin{figure}[ht]
\includegraphics[angle=0, width=80mm, height=75mm]{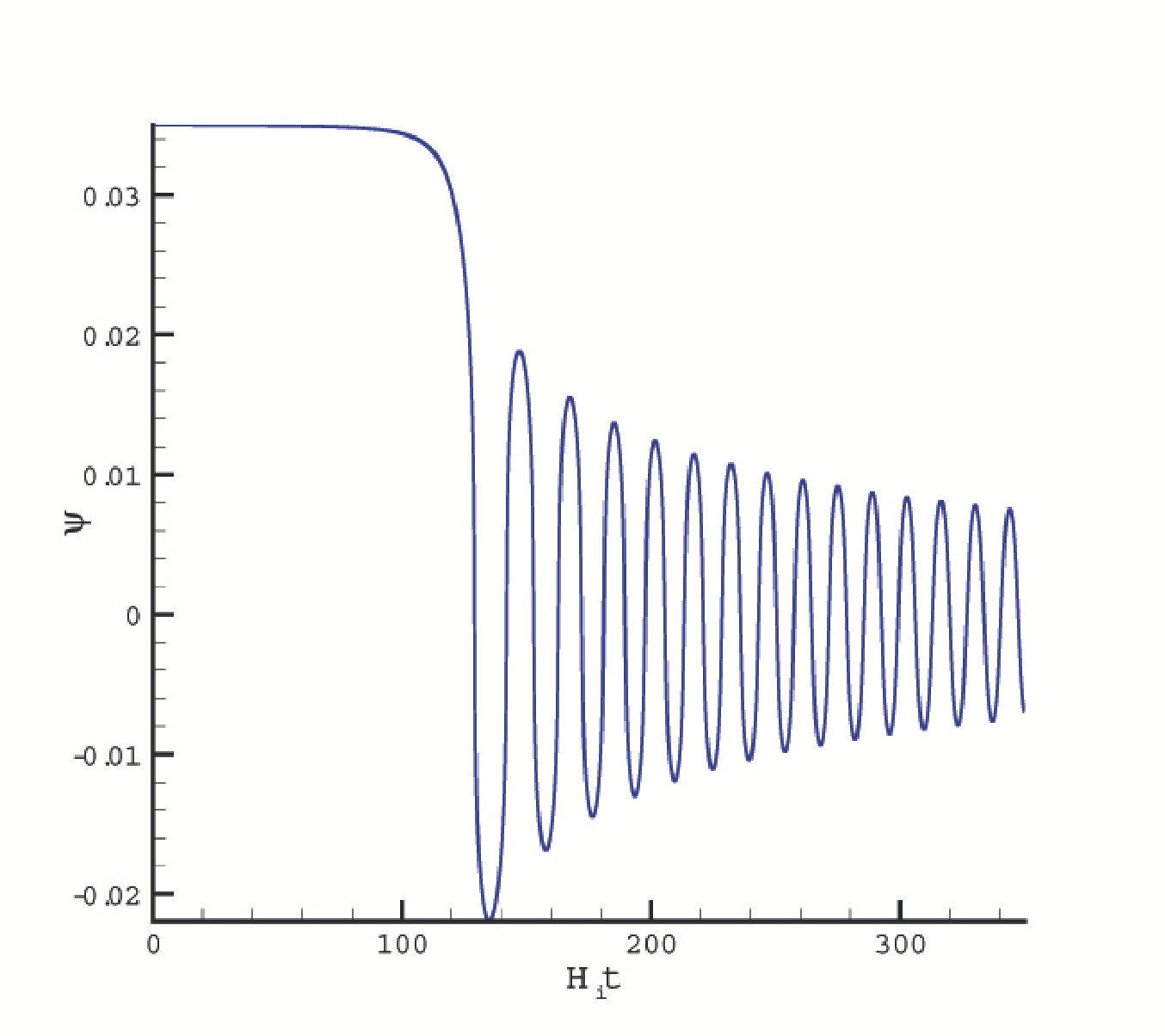}
\includegraphics[angle=0, width=80mm, height=75mm]{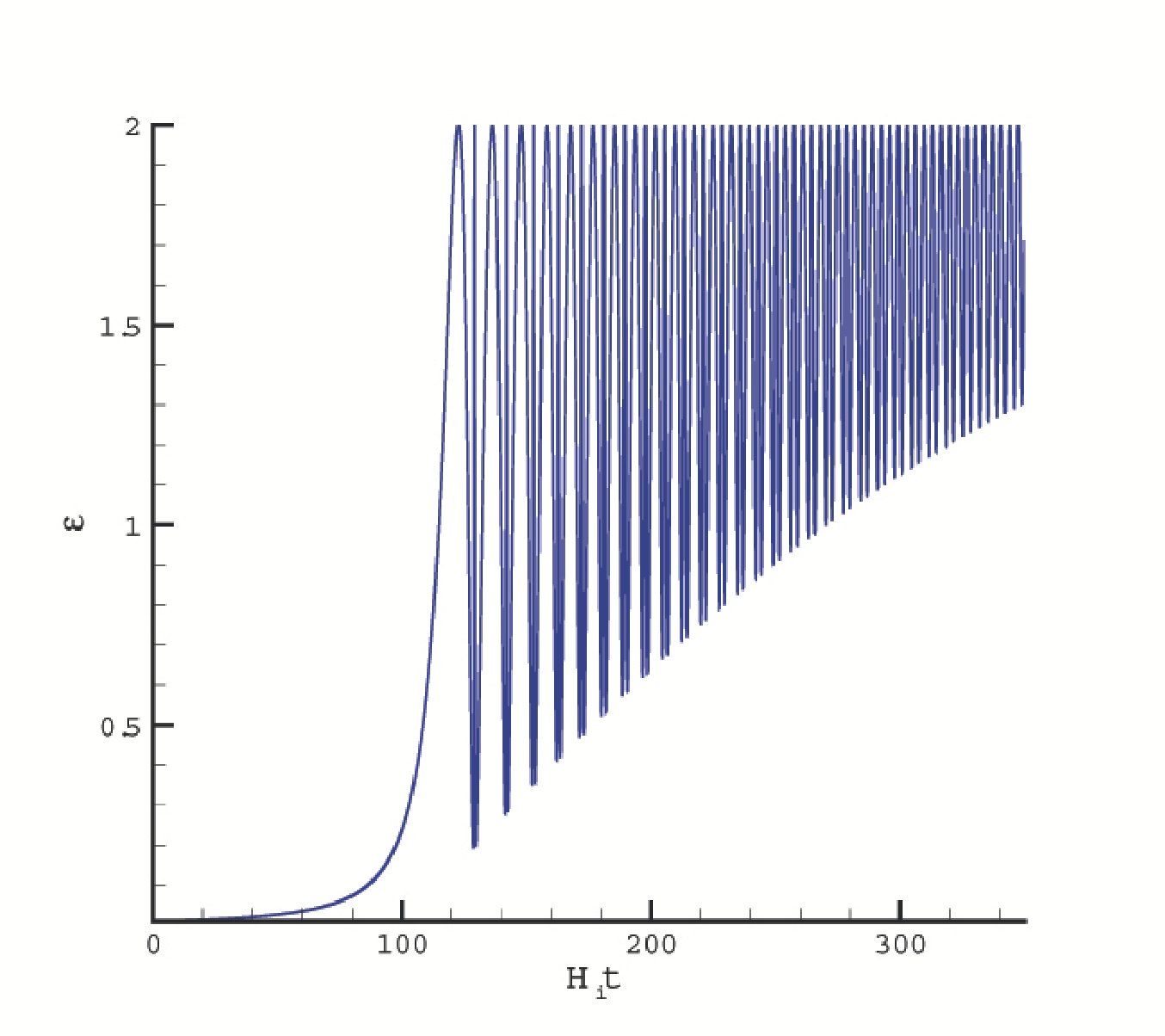}\\
\includegraphics[angle=0,width=80mm, height=75mm]{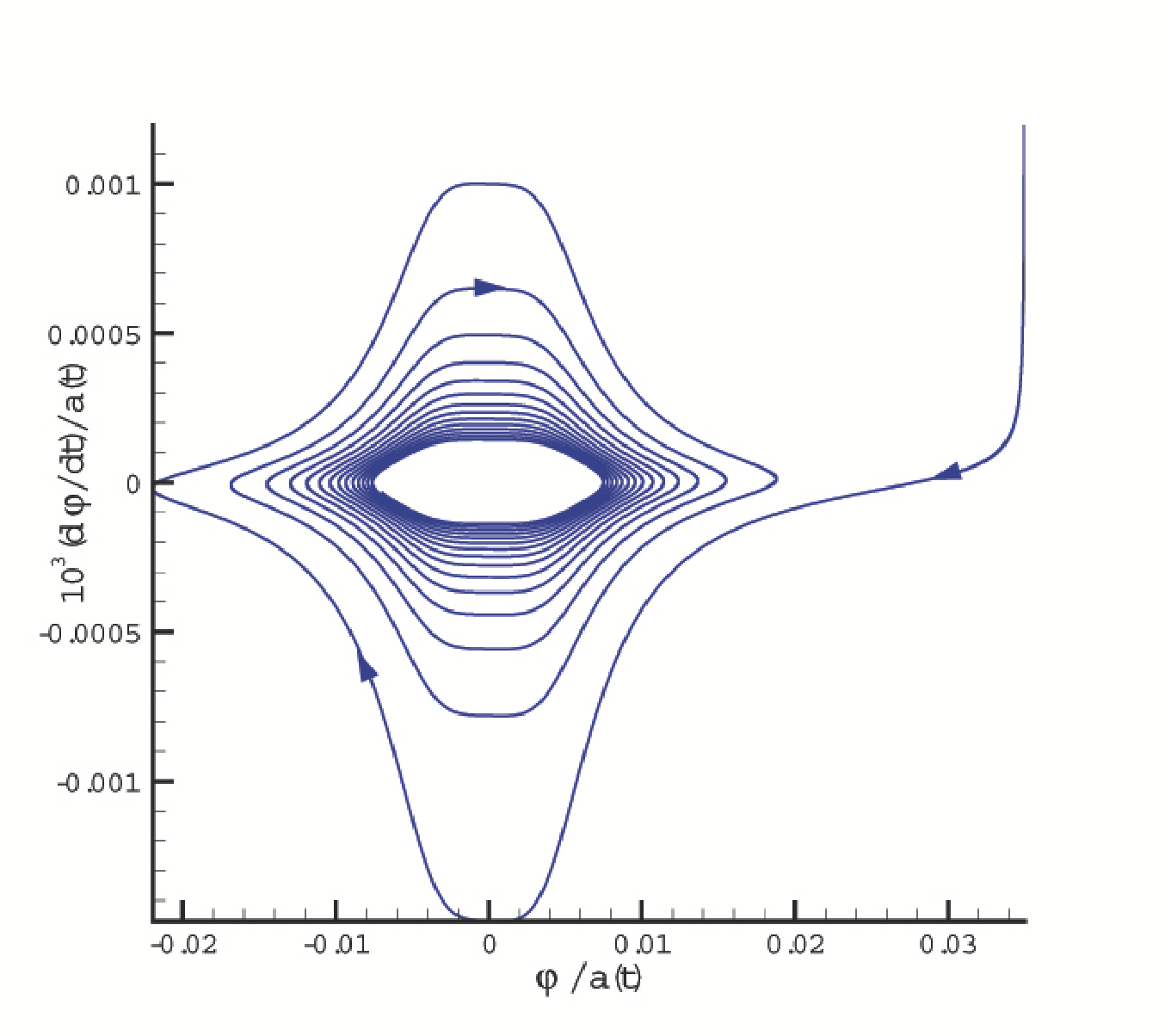}
\includegraphics[angle=0,width=80mm, height=75mm]{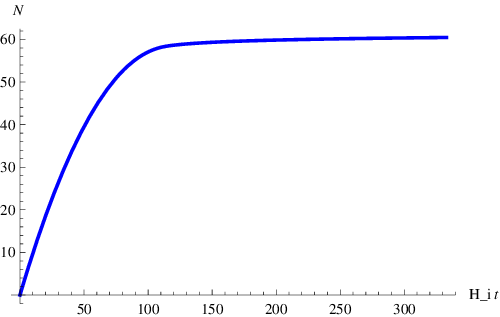}
\caption{The classical trajectory for ${\psi_i=0.035,\dot\psi_i=-10^{-10};\ g=2.5\times10^{-3}, \kappa=1.733\times 10^{14}}$. These values correspond to a slow-roll trajectory with $H_i=3.4\times 10^{-5},\ \gamma_i=6.62,\ \epsilon_i=9.3\times 10^{-3},\ \delta_i=8.4\times 10^{-5}$. These are the values very close to the range for which the gauge-flation is compatible with the current cosmological and CMB data (\emph{cf.} discussions of section \ref{testing-the-model}). Note that $\kappa,\ H_i$ and $\psi_i$ are given in the units of $\mpl$.}\label{x=6.35-slow-roll-figures}
\end{figure} 

This behavior is of course expected, noting \eqref{rho0-rho1} and that in the oscillatory regime the dominant term is the $\rho_{_{YM}}$, that is, the system effectively  behaves as a $g^2\psi^4$ chaotic inflation theory.
And it is well-known that after the slow-roll phase $\psi(t)$ in the $g^2\psi^4$ theory oscillates  as a Jacobi-cosine function whose amplitude drops like $t^{-1/2}$ \cite{preheating-Linde}.
In other words, the averaged value of $\epsilon$ and $a(t)$ behave like a radiation dominated Universe (recall that for a radiation dominated cosmology $\epsilon=-\dot H/H^2=2$).

\vskip 2mm
\noindent
$\blacktriangleright$ \emph{\textbf{Discussion on diagrams in Fig.\ref{x=6.35-slow-roll-figures}}.}

The top left figure shows evolution of the  effective inflaton field $\psi$ as a function of $H_it$. As we see, there is a period of slow-roll, where $\psi$ remains almost constant and $\epsilon$ is almost constant and very small. Toward the end of the slow-roll $\epsilon$ grows and becomes one (the top right figure), (slow-roll) inflation ends and $\psi$ suddenly falls off  and starts oscillating. As we see from the top right figure, the slow-roll parameter $\epsilon$ has an upper limit which is equal to 2. This is understandable recalling \eqref{epsilon-rho0-rho1} and that $\rho_\kappa$ is positive definite. At the end of slow-roll inflation  $\rho_\kappa$, is negligible and the system is essentially governed by the Yang-Mills part $\rho_{_{YM}}$. In addition, the top left figure shows that  amplitude of the
$\psi$ field in the oscillatory part is dropping like $t^{-1/2}$. (The  minima of $\epsilon$ is also  fit by a  $t^{1/2}$ curve.)

The bottom left figure shows the phase diagram of the effective inflaton trajectory. Note that this diagram depicts $\dot\phi/a(t)$ vs $\phi/a(t)$ (rather than $\dot\psi$ vs $\psi$).
The rightmost vertical line is where we have slow-roll, because $\phi=a(t)\psi$ and  during slow-roll $\psi$ is almost a constant. The curled up part is when inflation has ended, and when the system oscillates around a radiation dominated phase.
This latter may be seen in the figure noting that  the amplitudes
of oscillations of both $\dot\phi/a(t)$ and $\phi/a(t)$ drop by $t^{-1/2}$. The bottom right figure shows number of e-folds as a function of comoving time. As expected, the number of e-folds reaches its asymptotic value when $\epsilon\simeq 1$.

 One can readily check that the behavior of $\psi, \epsilon$, and the number of e-folds during slow-roll inflationary period has a perfect matching with our analytic results of previous section. We note that, as will be discussed  in section \ref{testing-the-model}, the set of parameters $H_i, \gamma_i, \psi_i, g$ corresponds to an inflationary model close to the range of values compatible with the current cosmological and CMB data.

\begin{figure}
\includegraphics[angle=0, width=80mm, height=80mm]{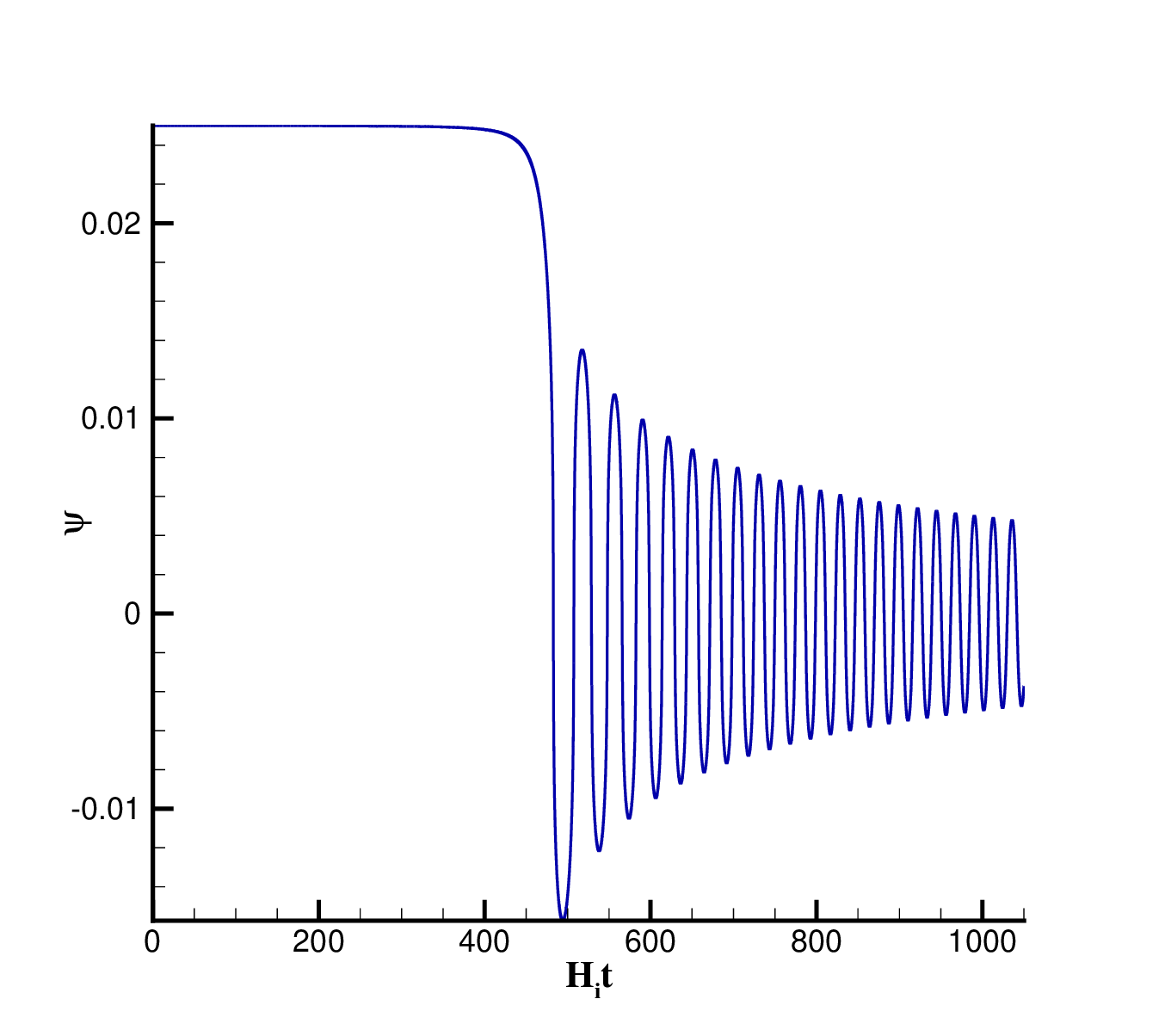}
\includegraphics[angle=0, width=80mm, height=80mm]{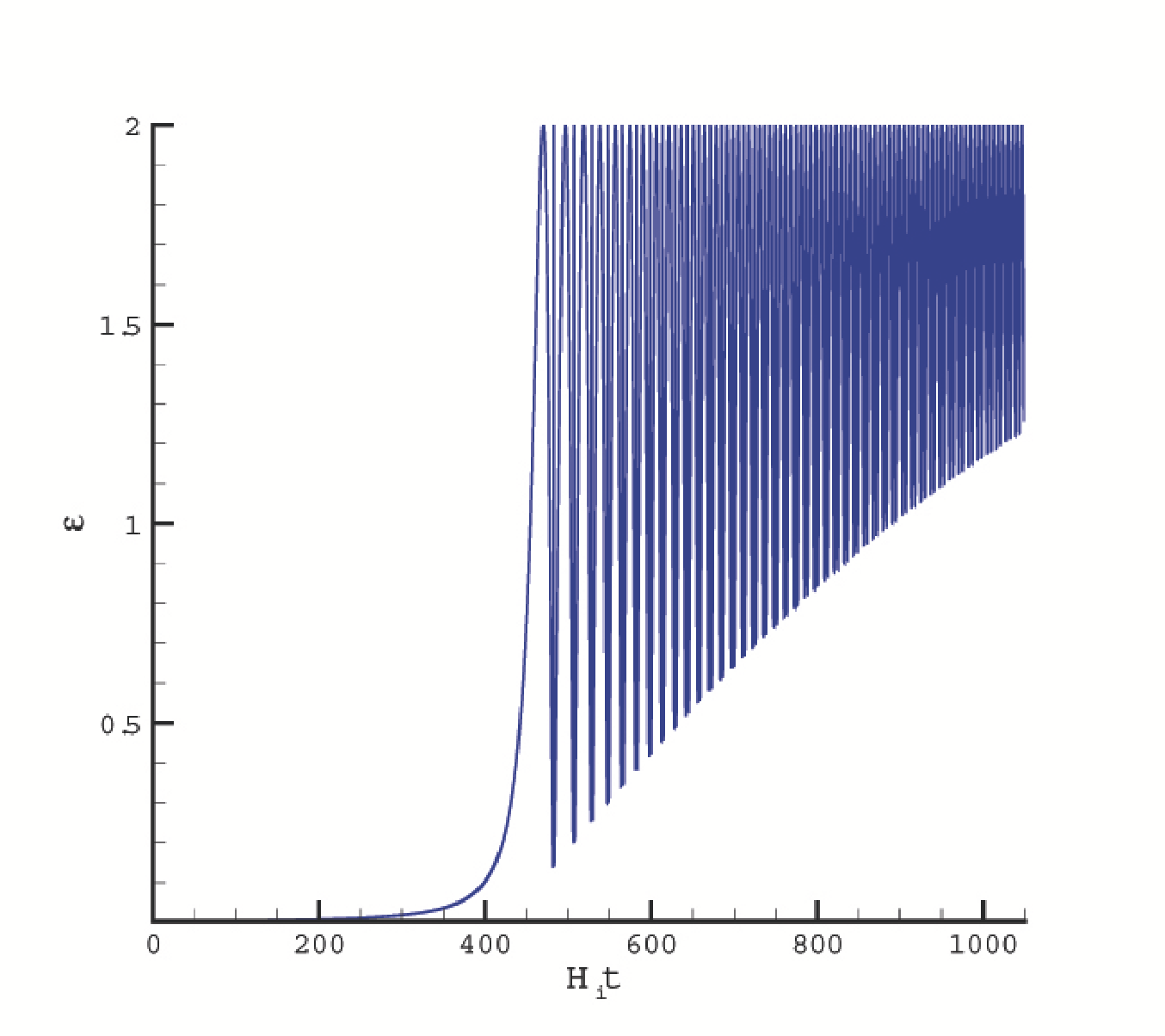}\\
\includegraphics[angle=0,width=80mm, height=80mm]{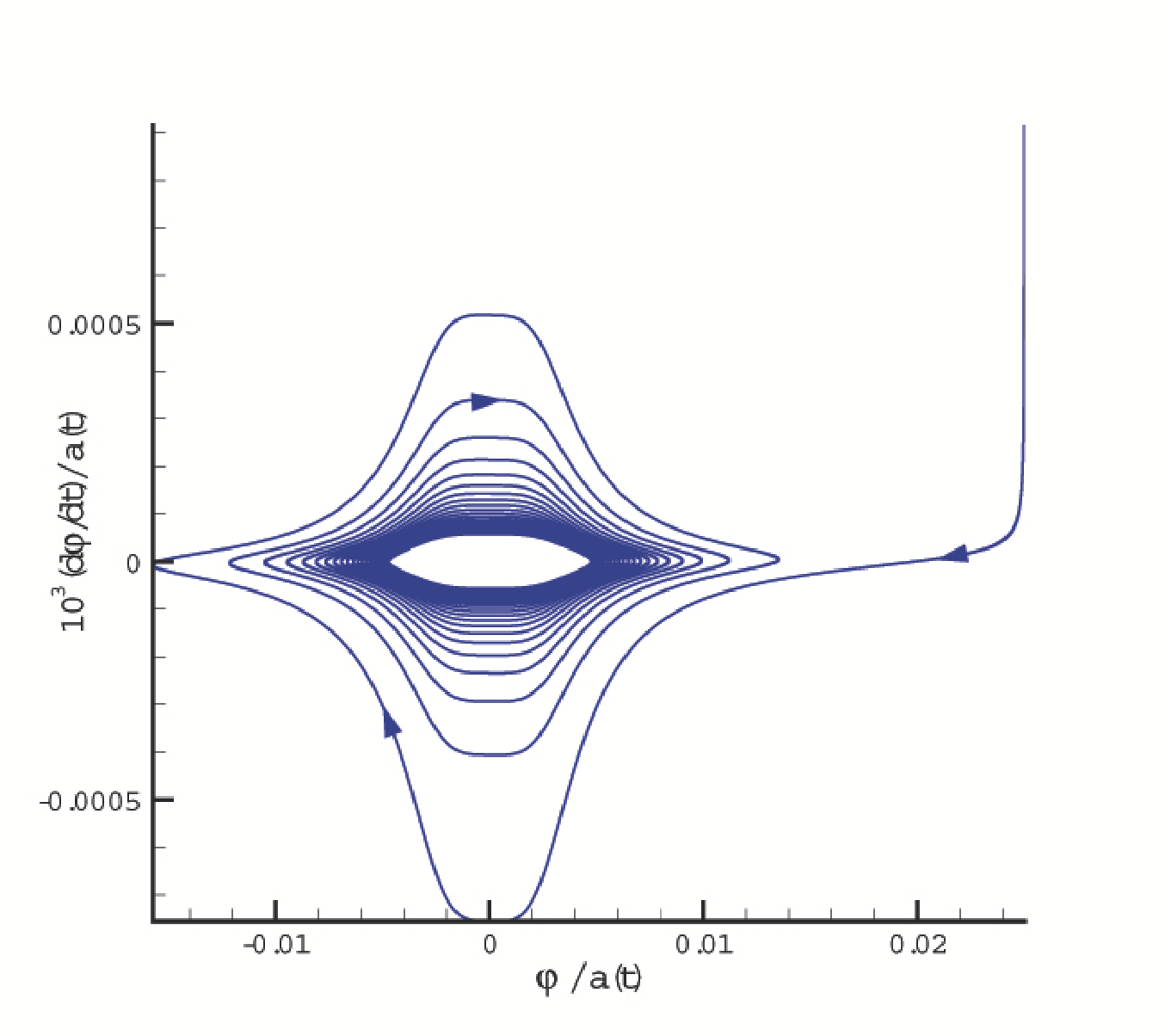}
\includegraphics[angle=0,width=70mm, height=70mm]{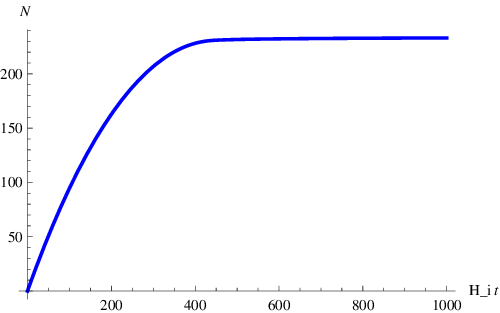}
\caption{The classical trajectory for ${\psi_i=0.025,\dot\psi_i=-10^{-10};\ g=2.507\times10^{-3}, \kappa=1.3\times 10^{15}}$. These values correspond to a \emph{slow-roll} trajectory with $H_i=3.63\times 10^{-5},\ \gamma_i=2.98,\ \epsilon_i=2.5\times 10^{-3},\ \delta_i=1.1\times 10^{-4}$.  These figures show that it is possible to get arbitrarily large numbers of e-folds within the \emph{slow-roll} phase of our gauge-flation model. 
}\label{N200-slow-roll-figures}\end{figure}

\begin{figure}
\includegraphics[angle=0, width=80mm, height=80mm]{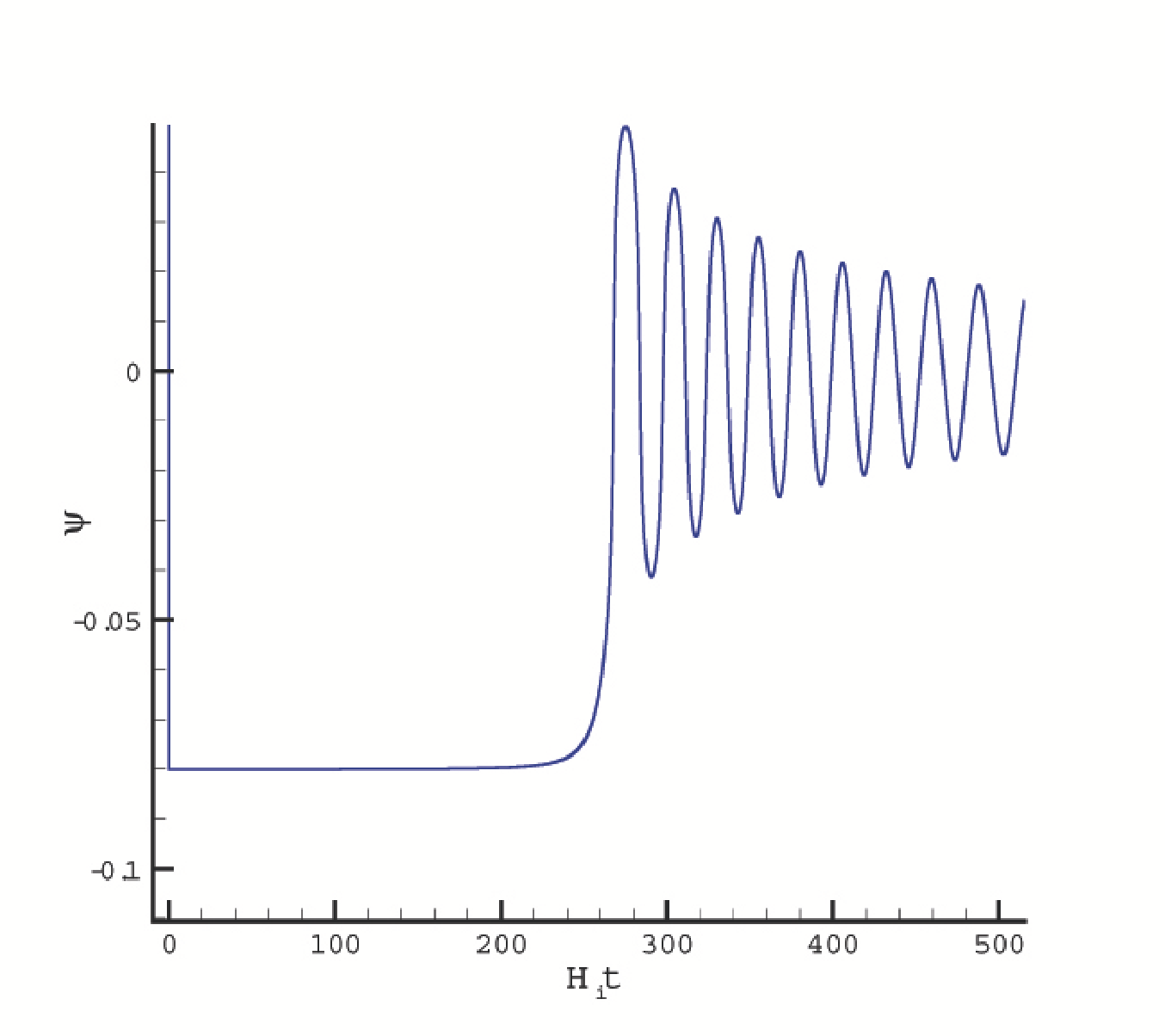}
\includegraphics[angle=0, width=80mm, height=80mm]{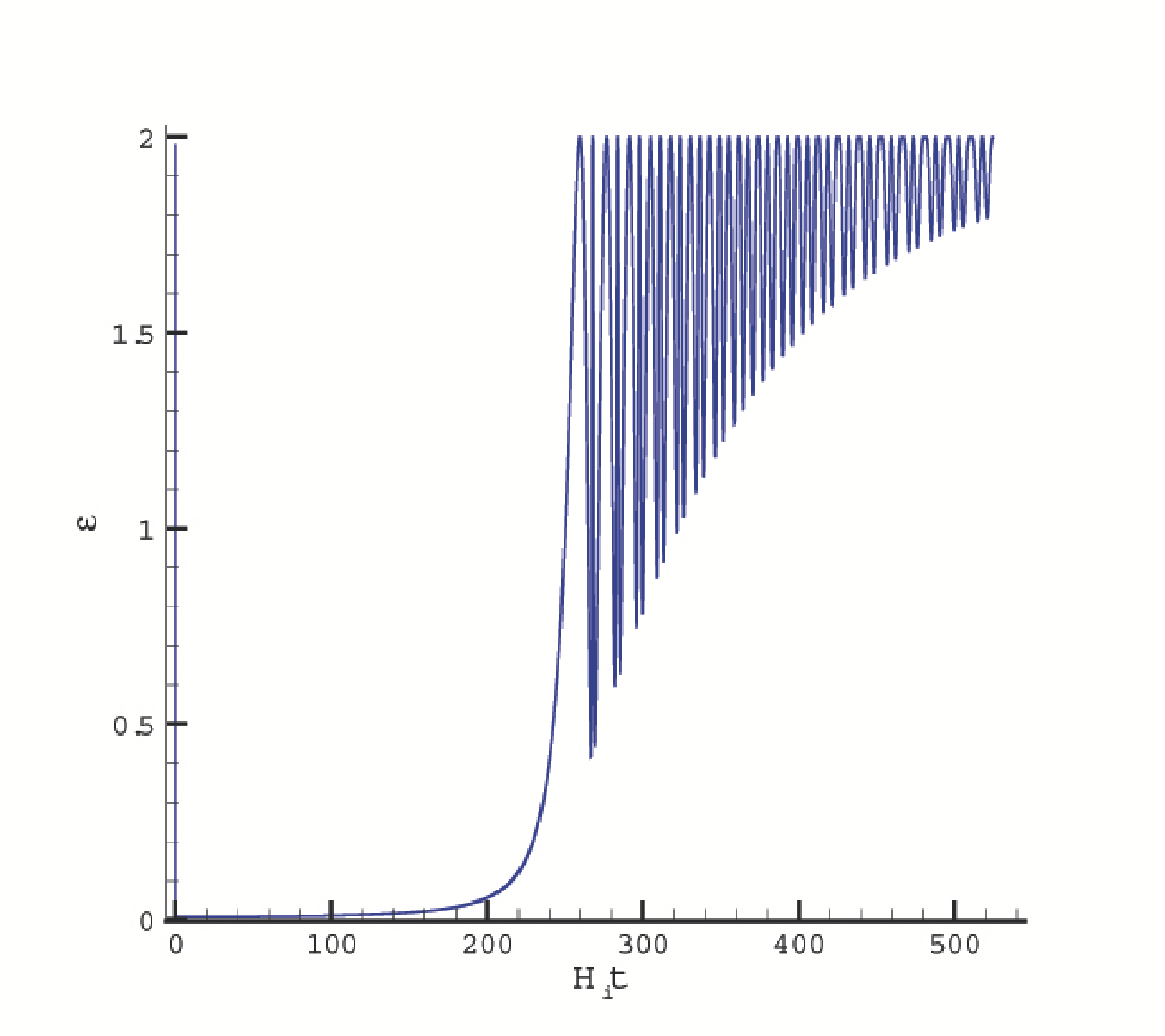}\\
\includegraphics[angle=0,width=80mm, height=80mm]{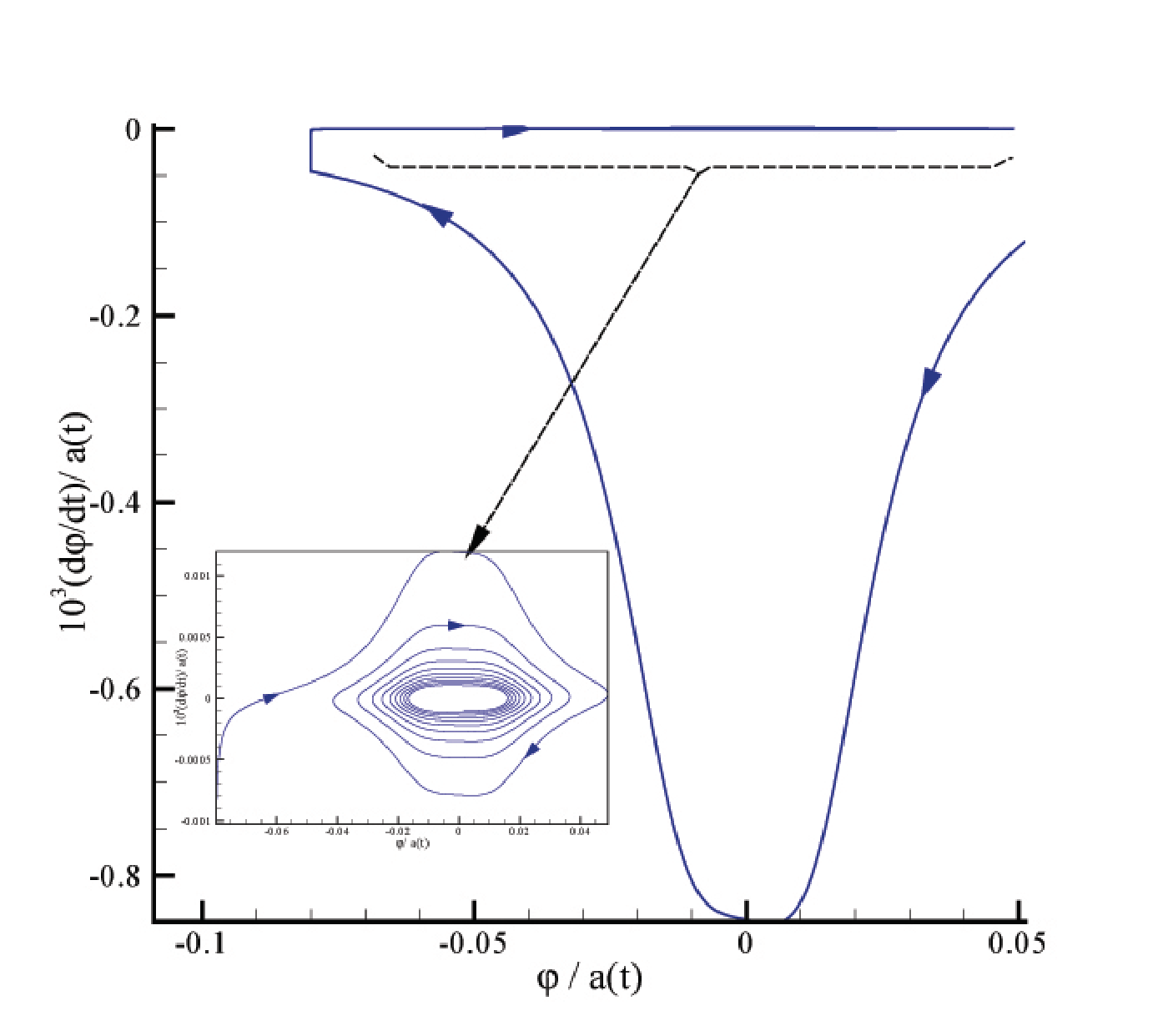}
\includegraphics[angle=0,width=70mm, height=70mm]{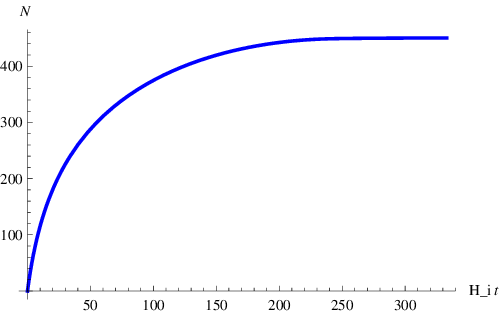}
\caption{Classical trajectory for ${\psi_i=8.0\times 10^{-2} ,\dot\psi_i=-10^{-4};\, g=4.004\times 10^{-4}\ , \kappa=4.73\times 10^{13}}$. These values correspond to a \emph{non-slow-roll} trajectory with $\delta\sim 2$, $H_i=6.25\times 10^{-4},\  \epsilon_i=6.4\times 10^{-3}$.
We start far from the slow-roll regime for which $\delta\sim \epsilon^2\ll 1$. This latter is also seen from the phase diagram (bottom left figure). Despite starting far from slow-roll regime, as we see from the top left figure, after an abrupt oscillation the field $\psi$ loses its momentum and falls into the standard slow-roll trajectory. As shown in the bottom right figure, for this case we get a large number of e-folds. Getting a large enough number of e-folds seems to be a fairly robust result not depending much on the initial value of $\delta$.
}\label{delta20-figures}\end{figure}

\vskip 2mm
\noindent
$\blacktriangleright$ \emph{\textbf{Discussions on diagrams in Figs.~\ref{N200-slow-roll-figures} and \ref{delta20-figures}}.}

Fig.~\ref{N200-slow-roll-figures} corresponds to a slow-roll trajectory which starts with a lower value of
$\epsilon$, but almost the same value for $H$, compared to the case of Fig.~\ref{x=6.35-slow-roll-figures}. For this case we hence get a larger number of e-folds. The qualitative shape of all four figures is essentially the same as those of Fig.~\ref{x=6.35-slow-roll-figures}, and both are compatible with our analytic slow-roll results of previous section.
Our numeric analysis indicates that the behavior of the phase diagram for $\psi$ field and $\epsilon$ do not change dramatically when the orders of magnitude of the initial parameters are within the range given in Figs.~\ref{x=6.35-slow-roll-figures} or \ref{N200-slow-roll-figures}.

Fig.~\ref{delta20-figures} shows a trajectory with a relatively large $\dot\psi$. As we see after a single fast falloff the field falls into usual slow-roll tracks, similar to what we see in Figs.~\ref{x=6.35-slow-roll-figures} and \ref{N200-slow-roll-figures}. The oscillatory behavior after the inflationary phase, too, is the same as those of slow-roll inflation. The graph in the square in the bottom left figure shows, with a higher resolution, the upper part of the phase diagram
which comes with under brace. This part corresponds to the dynamics of $\psi$ field after inflation ends, and as expected has the same qualitative form as the phase diagrams in the slow-roll trajectories  of Figs.~\ref{x=6.35-slow-roll-figures} and \ref{N200-slow-roll-figures}.
Our numeric analysis shows that getting a large enough number of e-folds and the generic after-inflation behavior of the fields is robust and does not crucially depend on the initial value of the fields, $\psi_i$ or $\dot\psi_i$, but it is sensitive to the initial value of $\epsilon$. More precisely, as long as $\epsilon$ remains small of order $0.01$, regardless of the value of $\delta$ we can get arbitrarily large number of e-foldings. The examples of small and large $\delta$ values have been, respectively,  given in Figs.~\ref{N200-slow-roll-figures} and \ref{delta20-figures}.

It is also useful to work out the displacement of the $\psi$ field during inflation. To this end, let us start from $\dot \psi=-\delta H\psi$ and use the value for $\delta$ given in \eqref{eta-x}. Our numerical analysis reveals that
the dynamics of the system is such that even if we start with $\frac{\dot\delta}{H\delta}$  of order $1-10$, but $\psi^2_i\sim\epsilon_i$, after a short time $H_it\sim 1$, $\dot\delta/(H\delta)$ becomes very small and hence in almost all the inflationary period, except for the first one or two e-folds, $\delta\simeq \frac{\gamma}{6(\gamma+1)}\epsilon^2$. Since variations of $\delta$ in this period happens very fast, and after that it remains almost zero, during this period $\psi$ does not change much. Numerical analysis also shows that these arguments are generically true even if we start with a large value of $\delta$, as in Fig.~\ref{delta20-figures}, provided that the other parameters are such that we get large number of e-folds (about $N_e\sim 60$ or larger). This latter can be easily arranged for. This is again confirming the robustness of the classical inflationary trajectories with respect to the choice of the initial conditions.
Therefore, we may confidently compute roaming of $\psi$ field using the equation
$$
\frac{\dot\psi}{\psi^5}\simeq \frac{g^2}{6} \frac{\dot H}{H^3}\,.
$$
Integrating the above equation one can
compute the displacement of the scalar field $\psi$ during inflation. If we denote the value of $\psi$ in the beginning and end of inflation, respectively, by $\psi_i$ and $\psi_f$ we obtain%
\be%
\psi_f^2\simeq \frac{\sqrt3}{2}\ \psi_i^2\,.
\ee%
Alternatively one could have used \eqref{kappa-x} $\kappa g^2\psi^6(1-\delta)^2=2-\epsilon$ to compute the change in $\psi$. If we are in slow-roll regime where we can drop $\delta$-term, then the ratio of $\psi_f$ (which is computed for $\epsilon_f=1$) to $\psi_i$ is obtained as $\psi_f^6=\frac12\psi_i^6$. And up to percent level, $\sqrt{3}/2$ and $2^{-1/3}$ are equal, a confirmation of the validity of the slow-roll approximations we have used.
Interestingly, at this level of approximation, the roaming of the $\psi$ field is independent of the initial value of the $\gamma$ (or the initial value of Hubble) parameter and $\psi_i$ and $\psi_f$ have the same orders of magnitude.

\section{Cosmic perturbation theory in gauge-flation}\label{GF-cosmic-pert-theory-section}

So far we have shown that our gauge-flation model can produce a fairly standard slow-roll inflating Universe
with enough number of e-folds. The main test of any inflationary model, however, appears in the  imprints inflation has left on the CMB data, \ie, the power spectrum of curvature perturbations and primordial gravity waves, and the spectral tilt of these spectra. To this end, we should go beyond the homogeneous ($x$-independent) background fields and consider fluctuations around the background. This is what we will carry out in this section.

\subsection{Gauge-invariant metric and gauge field perturbations}

Although not turned on in the background, all of the components of  metric and the gauge field in all gauge and spacetime directions will have quantum fluctuations and should be considered. The metric perturbations may be parameterized in the standard form \cite{Inflation-Books,Bassett-review}
\be\label{metric-pert}%
\begin{split}
ds^2&=-(1+2A)dt^2+2a(\partial_iB-S_i)dx^idt\\
&+a^2\left((1-2C)\delta_{ij}+2\partial_{ij}E+2\partial_{(i}W_{j)}+h_{ij}\right)dx^idx^j\,,
\end{split}
\ee
where $\partial_i$ denotes partial derivative respect to $x^i$ and $A,\ B,\ C$ and $E$ are scalar perturbations, $S_i,\ W_i$ parameterize vector perturbations (these are divergence-free three-vectors) and $h_{ij}$, which is symmetric, traceless and divergence-free, is the tensor mode.
The 12 components of the gauge field fluctuations may be parameterized as%
\bea \label{gauge-field-pert-1}
\delta A^a_{~0}&=&\delta^{k}_a\partial_k\dot{Y}+\delta_a^j u_j\,,\\  \label{gauge-field-pert-2}
\delta A^a_{~i}&=&\delta^a_i Q+\delta^{ak}\partial_{ik}M+g\phi\epsilon^{a~k}_{~i}\partial_{k}P+\delta_a^j\partial_i v_j+\epsilon^{a~j}_{~i}w_j+\delta^{aj}t_{ij}\,,
\eea
where, as discussed in section \ref{setup-section}, we have identified the gauge indices with the local Lorentz indices and the expansion is done around the background in $A_0^a=0$ temporal gauge. In the above, as has been made explicit, there are four scalar perturbations, $Q,\ Y,\ M$ and $P$, three divergence-free three-vectors $u_i,\ v_i$ and $w_i$, and a symmetric traceless divergence-free tensor $t_{ij}$, adding up to $4+3\times 2+2=12$.
$Q$ is the perturbation of the background field $\phi$, which is the only scalar in the perturbed gauge field without spacial derivative.
We are hence dealing with a situation similar to the multifield inflationary theories where we have adiabatic and isocurvature perturbations. If the analogy held, $Q$ would have then be like the adiabatic mode. However, as we will see this is not true and the curvature perturbations are dominated by other scalars and not $Q$. As another peculiar and specific feature of the gauge-flation cosmic perturbation theory, not shared by any other scalar-driven inflationary model, we note that the gauge field fluctuations contain a tensor mode $t_{ij}$.

Because of the presence of gauge symmetries not all $10+12$ metric plus gauge field perturbations are physical. Altogether there are four diffeomorphisms and three local gauge symmetries, hence we have $15$ physical degrees of freedom. The four diffeomorphisms remove two scalars and a divergence-free vector \cite{Inflation-Books}, and the three gauge transformations one scalar and one divergence-free vector. Therefore, we have \emph{five} physical  \emph{scalar} perturbations, \emph{three} physical divergence-free \emph{vector}, and \emph{two} physical \emph{tensor} perturbations. (These amount to $5\times 1+3\times 2+2\times 2=15$ physical degrees of freedom.)
The gauge degrees of freedom may be removed by gauge-fixing (working in a specific gauge) or working with gauge-invariant combinations of the perturbations. In what follows we work out the gauge- and diffeomorphism-invariant combinations of these modes.

\subsubsection{Scalar modes}

Let us first focus on the scalar perturbations   $A$, $B$, $C$, $E$, $Q$, $\dot{Y}$, $M$ and $P$.
Under infinitesimal scalar coordinate transformations
\be  \label{coor}
\begin{split}
t&\rightarrow \tilde{t}=t+\delta t\,,\\
x^i&\rightarrow \tilde x^i=x^i+\delta^{ij}\partial_j \delta x\,,
\end{split}
\ee
where $\delta t$ determines the time slicing and $\delta x$ the spatial threading,
the scalar fluctuations of the gauge field and  metric transform as
\be
\begin{split}
Q\rightarrow Q-\dot{\phi}\delta t\,&,\qquad
\dot{Y}\rightarrow\dot{Y}-\phi\dot{\delta x}\,,\\
M\rightarrow M-\phi\delta x\,&,\qquad
P\rightarrow P\,,\\
A \rightarrow A-\dot{\delta t}\,&,\qquad
C\rightarrow C+H\delta t\,,\\
B \rightarrow B+\frac{\delta t}{a}-a\dot{\delta x}\,&,\qquad
E\rightarrow E-\delta x\,.
\end{split}
\ee

On the other hand  under an infinitesimal gauge transformation $\lambda^a$, fluctuations of the gauge field transform as%
\be \label{gua}
\delta A^a_{~\mu}\rightarrow \delta A^a_{~\mu}-\frac{1}{g}\partial_\mu\lambda^a-\epsilon^a_{~bc}\lambda^b A^c_{~\mu}\,.
\ee
The gauge parameters $\lambda^a$ can be decomposed into a scalar and a divergence-free vector:
\be
\lambda^a=\delta^{ai}\partial_i\lambda+\delta^a_i\lambda^{~i}_{V}\,.
\ee
The scalar part of the gauge field perturbations under the action of the scalar gauge transformation $\lambda$ transform as
\be\begin{split}
Q \rightarrow Q\,& ,\qquad
Y \rightarrow Y-\frac{1}{g}\lambda\,,\\
M\rightarrow M-\frac{1}{g}\lambda\,&,\qquad
P\rightarrow P+\frac{1}{g}\lambda\,.
\end{split}
\ee
We note that $Q$ is gauge-invariant and this is a result of identifying the gauge indices with the local Lorentz indices and that $Q$ is a scalar.

Equipped with the above, one may construct five independent gauge-invariant combinations. One such choice is \footnote{These choices are not unique and one can construct other gauge invariant combinations too.}
\bea
\Psi&=&C+a^2H(\dot{E}-\frac{B}{a})\,,\\
\Phi&=&A-\frac{d}{dt}\left(a^2(\dot{E}-\frac{B}{a})\right)\,,\\
\mathcal{Q}&=&Q+\frac{\dot\phi}{H}C\,,\\
\mathcal{M}&=&\frac{g^2\phi^3}{a^2}(M+P-\phi E)\,,\\
\tilde{\mathcal{M}}&=&\dot\phi(\dot M-\dot\phi E-\dot Y)\,.
\eea
The first two, $\Phi$ and $\Psi$ are the standard Bardeen potentials, while
$\mathcal{Q}$, $\mathcal{M}$ and $\tilde{\mathcal{M}}$ are the three gauge and diffeomorphism-invariant combinations coming from the gauge field fluctuations.

Finally, for the later use we also present the first-order perturbations of the gauge field strength sourced by the scalar perturbations%
\bea \label{Fmunu-scalar}
\delta F^a_{~0i}&=&\delta^a_i\dot{Q}+\delta^{aj}\partial_{ij}(\dot{M}-\dot{Y})-g\epsilon^{aj}_{~~i}\partial_j((\phi P\dot{)}+\phi\dot{Y})\,,\\
\delta F^a_{~ij}&=&2\delta^a_{[j}\partial_{i]}(Q+g^2\phi^2P)+2g\phi\epsilon^{ak}_{~~[i}\partial_{j]k}(P+M)-2g\epsilon^a_{~ij}Q\phi\,.
\eea
Note that $\delta F^a_{\mu\nu}$ are not gauge-invariant, as under gauge transformations $F^a_{\mu\nu}\to F^a_{\mu\nu}- \epsilon^a_{\ bc} \lambda^b F^c_{\mu\nu}$.

\subsubsection{Vector modes}

Next, we consider the vector modes $S_i$, $W_i$, $u_i$, $v_i$ and $w_i$.
Under   infinitesimal ``vector'' coordinate transformations
\be  \label{veccoor}
x^i\rightarrow \tilde x^i=x^i+\delta x_V^i\,,
\ee
where $\partial_i\delta x^V_i=0$,
\be
\begin{split}
S_i\rightarrow S_i+a\delta\dot{x}_V^i\,&,\qquad W_i \rightarrow W_i-\delta x_V^i\,,\\
u_i \rightarrow u_i-\phi\dot{\delta x}^i_V\,,\quad &v_i \rightarrow  v_i-\phi\delta x^i_V\,,\quad w_i\rightarrow w_i\,.
\end{split}
\ee
On the other hand  under the vector part of infinitesimal gauge transformation \eqref{gua},
\be
u_i \rightarrow u_i-\frac{1}{g}\dot{\lambda}_V^i\,,\quad
v_i \rightarrow v_i-\frac{1}{g}\lambda_V^i\,,\quad
w_i \rightarrow w_i+\phi\lambda_V^i\,,
\ee
and obviously $S_i,\ W_i$ remain invariant.

The three gauge- and diffeomorphism-invariant divergence-free vector perturbations may be identified as%
\bea
\mathcal{Z}_i&=&a\dot{W}_i+S_i\,,\\
\mathcal{U}_i&=&u_i-\phi \dot W_i+\frac{1}{g}\dot{w}_i\,,\\
\mathcal{V}_i&=&v_i-\phi W_i+\frac{1}{g}w_i\,.
\eea

The contribution of vector perturbations to the first-order gauge field strength perturbations are
\be \label{Fmunu-vector}
\begin{split}
\delta F^a_{~0i}=&\delta_a^j\partial_i(\dot{v}_j-u_j)
+\epsilon^{a~j}_{~i}(\dot{w}_j+g\phi u_j)\,,\\ 
\delta F^a_{~ij}=&2\epsilon^{a~~k}_{~[j}\partial_{i]}(w_k+g\phi v_k)+2g\phi\delta^a_{~\;[i}w_{j]}\,.
\end{split}\ee

\subsubsection{Tensor modes}

One can show that the tensor perturbations $h_{ij}$ and $t_{ij}$, being symmetric, traceless and divergence-free, are both gauge- and
 diffeomorphism-invariant. The contribution of $t_{ij}$ to the first order perturbed $F^a_{\mu\nu}$ corresponding to $t_{ij}$ is
\be\label{Fmunu-tensor}%
\begin{split}
\delta F^a_{0i}=&\delta^{aj}\dot{t}_{ij}\,,\\
\delta F^a_{0i}=&2\delta^{ak}\partial_{[i}t_{j]k}-2g\phi\epsilon^{ak}_{~~~[j}t_{i]k}\,.
\end{split}
\ee

\subsection{Field equations}

Having worked out the gauge invariant combinations of the field perturbations, we are now ready to study their dynamics. These first order perturbations are governed by perturbed Einstein and gauge field equations%
\be
\delta G_{\mu\nu}=\delta T_{\mu\nu}\,,\qquad \delta \left(D_\mu \frac{\partial\mathcal{L}}{\partial F^a_{\mu 0}}\right)=0\,,
\ee
where by $\delta$ in the above we mean first order in field perturbations. Note that the zero element of the gauge field equation (the equation of motion of $A^a_0$) is a constraint enforcing the gauge invariance of the action and hence independent of the Einstein equation.

Since we are dealing with an isotropic perfect fluid in the background, as it is customary in standard cosmology text books \cite{Inflation-Books}, it is useful to decompose energy-momentum perturbations as%
\begin{subequations}
\begin{align}
\delta T_{ij}=&P_0\delta g_{ij}+a^2\left(\delta_{ij}\delta P+\partial_{ij}\pi^s+\partial_i\pi^V_j+\partial_j\pi^V_i+\pi^T_{ij}\right)\,,\\
\delta T_{i0}=&P_0\delta g_{i0}-(P_0+\rho_0)(\partial_i\delta u+\delta u_i^V)\,,\\
\delta T_{00}=&-\rho_0\delta g_{00}+\delta \rho\,,
\end{align}
\end{subequations}
where subscript, $``0"$ denotes a background quantity and $\pi^s$,
$\pi^V_i$ ,~ $\pi^T_{ij}$ represent the anisotropic inertia and
characterize departures from the perfect fluid form of the
energy-momentum tensor; $\pi^V_i$ and $\pi^T_{ij}$ and  the
vorticity $\delta u^V_i$ satisfy \be
\partial_i\delta u^V_i=\partial_i\pi^V_i=0\,,\qquad \pi^T_{ii}=0\,,\qquad \partial_i\pi^T_{ij}=0\,.
\ee
Since being a perfect fluid or having irrotational flows are physical properties, their corresponding conditions are gauge-invariant.

\subsubsection{Scalar modes}

As is usually done in cosmic perturbation theory, it is useful to write down the equations of motion in a gauge-invariant form. In order this we note that $\delta T_{\mu\nu}$ has four gauge-invariant scalar parts
$\dre_g$, $\dpe_g$, $\dqe_g$,
\bea
\dre_g&=&\dre-\dot{\rho}_0a^2(\dot{E}-\frac{B}{a})\,,\\
\dpe_g&=&\dpe-\dot{P}_0a^2(\dot{E}-\frac{B}{a})\,,\\
\dqe_g&=&\dqe+(\rho_0+P_0)a^2(\dot{E}-\frac{B}{a})\,,
\eea
and $\pi^s$  \cite{Inflation-Books}, where $\delta q=(\rho_0+P_0)\delta u$.

After lengthy calculations (confirmed by \texttt{Maple} codes too), we obtain
\bea\label{PiS}
a^2\pi^S&=&2(\mM-\tM),\\
\dqe_g&=&-2(\dot \mM+3H\delta\times \mM-H\mM-\frac{g^2\phi^3}{\dot{\phi}a^2}\tM+\frac{\dot{\phi}}{a}(\frac{\mQ}{a}-\frac{\dot\phi}{aH}\Psi)),\\
\dre_g&=&3(1+\kk)\frac{\dot \phi}{a^2}(\dot \mQ-\frac{\dot\phi}{H}\dot\Psi)+6(1+\dk)\frac{g^2\phi^3}{a^3}\frac{\mQ}{a}-3(1+\kk)\frac{\dot\phi^2}{a^2}\Phi\nonumber\\
&+&3\epsilon\YM2\Psi-(1+\kk)\frac{k^2}{a^2}\tM-2(1+\dk)\frac{k^2}{a^2}\mM,\\
\dpe_g&=&(1-3\kk)\frac{\dot \phi}{a^2}(\dot \mQ-\frac{\dot\phi}{H}\dot\Psi)+2(1-3\dk)\frac{g^2\phi^3}{a^3}\frac{\mQ}{a}-(1-3\kk)\frac{\dot\phi^2}{a^2}\Phi\nonumber\\
&-&(4\frac{\dot\phi^2}{a^2}+3\frac{g^2\phi^4}{a^4})\epsilon\Psi-(\frac13-\kk)\frac{k^2}{a^2}\tM-2(\frac13-\dk)\frac{k^2}{a^2}\mM.
\eea
We are now ready to write down the four independent perturbed scalar Einstein equations, three of which are constraints and one is dynamical:%
\bea
\label{pi-S} &~&a^2\pi^S=\Psi-\Phi\,,\\
\label{dq} &~&\dqe_g+2(\dot{\Psi}+H\Psi)=0\,,\\
\label{drho} &~&\dre_g-3H\dqe_g+2\frac{k^2}{a^2}\Psi=0\,,\\
\label{dP} &~&\dpe_g+\dot{\dqe}_g+3H\dqe_g+(\rho_0+P_0)\Psi=0\,.
\eea
Although it is not independent of the Einstein equations, here for later convenience we also write the equation of
energy conservation
\be\label{energy-conservation}
\delta\dot\rho_g-3H\delta\dot q_g+3\epsilon H^2\delta q_g-6H\frac{k^2}{a^2}\Psi+H\frac{k^2}{a^2}(\Psi-\Phi)+2\frac{k^2}{a^2}(\dot\Psi+H\Phi)=0.
\ee
The above perturbed scalar Einstein equations do not suffice to deal with five gauge-invariant scalar degrees of freedom and one equation is missing. This last equation is, of course, provided by the perturbed gauge field equations of motion (constraint equation $D_\mu\big(\frac{\partial\cal L}{\partial F^a_{~0\mu}}\big)=0$). After using \eqref{drho}, this equation reads as\footnote{The constraint \eqref{A0eq} is equal to the equation of motion of $\tM$ from the second order action for the scalar perturbations \eqref{scalar-2nd-action}}
\be\label{A0eq}
\epsilon H^2\frac{Q}{\phi}+\delta\times H(\dot\Psi+H\Phi)-\frac{1}{2}\YM2\epsilon\Psi+\frac16\frac{k^2}{a^2}(\Psi+\Phi)=0.
\ee

These relations (\eqref{pi-S}-\eqref{A0eq}) provide enough number of equations for the
gauge-invariant scalar perturbations to which we return in the next subsection.

\subsubsection{Vector modes}

To study the vector perturbations, we first work out vector parts of
the perturbed energy-momentum tensor, $\delta q_i^V$ and
$\pi^V_i$, using \eqref{Fmunu-vector}:\footnote{Note that at first order in perturbation theory only the vector perturbations contribute to the vector part of energy-momentum tensor perturbations.}%
\bea\label{deltaq}
\delta q^V_i&=&-2\frac{g^2\phi^3}{a^2}\left(\mathcal{U}_i+\frac{\phi}{a}\mathcal{Z}_i\right)+\frac{g\phi^2}{a^2}
\left(\nabla\times(\dot{\vec{\mathcal{V}}}-\vec{\mathcal{U}})\right)_i
-\frac{g\phi\dot{\phi}}{a^2}\left(\nabla\times\vec{\mathcal{V}}\right)_i\,,\\
\label{piv}a\pi_i^V&=&\frac{g^2\phi^3}{a^3}\mathcal{V}_i
+\frac{\dot{\phi}}{a}(\mathcal{U}_i-\dot{\mathcal{V}}_i)\,. \eea
The perturbed Einstein equations have two vector equations, one constraint and one dynamical equation. These equations are%
\bea \label{firstV}
&~&\partial_i\left(2a^2\pi^V_j-\frac{1}{a}(a^2\mathcal{Z}_j\dot{)}\right)=0\,,\\\label{delta-q-V}
&~&2a\delta q_i^V+\nabla^2\mathcal{Z}_i=0\,.
\eea
In order to fully determine the system, we need one more equation, provided by the gauge field equation $(D_\mu\frac{\partial\mathcal{L}}{\partial F^a_{\mu0}}=0)$.
This constraint enforces that the momentum
conjugate to $u_i$ is vanishing and yields to
\be\label{2nd-vector}
-2\frac{g^2\phi^3}{a^2}(\mathcal{U}_i+\frac{\phi}{a}\mathcal{Z}_i)
+\frac{g\phi^2}{a^2}\big(\vec\nabla\times(\dot{\vec{\mathcal{V}}}
+\frac{\phi}{a}\vec{\mathcal{Z}})\big)_i-\frac{g\phi\dot\phi}{a^2}(\vec\nabla\times\vec{\mathcal{V}})_i
-\frac{\phi}{a^2}\nabla^2(\mathcal{U}_i-\dot{\mathcal{V}}_i)=0.
\ee
Using \eqref{delta-q-V}, the above equations leads to the following simple equation
\be\label{vec-const}
\frac{g\phi^2}{a^2}\big(\vec\nabla\times(\vec{\mathcal{U}}+\frac{\phi}{a}\vec{\mathcal{Z}})\big)_i-
\frac{\phi}{a^2}\nabla^2(\mathcal{U}_i-\dot{\mathcal{V}}_i)+\frac{1}{2a}\nabla^2{\cal Z}_i=0 \,,
\ee
which completes the set of equations we need for solving vector perturbations.
Combining \eqref{firstV}-\eqref{delta-q-V} and \eqref{vec-const}, we learn that $\cal Z$ is damping exponentially during the inflation.
Then, equations \eqref{deltaq} and \eqref{firstV} indicate that  $\mathcal{Z}_i$ vanishes
after horizon crossing.

To summarize, similar to the usual scalar-driven inflationary models, in gauge-flation the vector modes are
diluted away by the (exponential) accelerated expansion of the Universe during inflation. In other words, despite
of having vector gauge fields as inflaton, in our model vector modes are unimportant in inflationary cosmology.

\subsubsection{Tensor modes}

As discussed, there are two gauge- and diffeomorphism-invariant tensor modes $h_{ij}$ and $t_{ij}$,
while perturbed Einstein equations only lead to one equation for $h_{ij}$.
This equation, which is sourced by the contribution of $t_{ij}$ to the energy-momentum tensor, reads as%
\be \label{T}
\ddot{h}_{ij}+3H\dot{h}_{ij}+\frac{k^2}{a^2}h_{ij}=2\pi^T_{ij}\,.
\ee The other equation of motion is provided with the perturbed
gauge field equations of motion. After a tedious but straightforward
calculation, which is also confirmed by the \texttt{Maple} codes, we
obtain the following second-order action for the tensor modes 
\begin{align}
\label{2ndts}
\delta S^{(2)}_T\simeq\frac{1}{2}\int d^3xdt  a^3&\biggl[\frac{1}{2}(\frac{\dot{\phi}^2}{a^2}-\frac{g^2\phi^4}{a^4})h_{ij}^2+(-2\frac{\dot{\phi}}
{a}\frac{\dot{t}_{ij}}{a}+2\frac{g^2\phi^3}{a^3}\frac{t_{ij}}{a})h_{ij}
+\frac{1}{4}(\dot{h}_{ij}^2-\frac{k^2}{a^2}h^2_{ij})\cr
&+\frac{1}{a^2}\bigg(\dot{t}^2_{ij}-\frac{k^2}{a^2}t^2_{ij}-\epsilon\frac{\kappa g^2\phi^2}{a^2}\frac{\dot{\phi}^2}{a^2}t^2_{ij}\bigg)+\\ &+\bigg((\frac{\kappa g\phi^2\dot\phi}{a^3}\dot{)}-2\frac{g\phi}{a}\bigg)\epsilon^{ijk}\frac{1}{a^3}t_{kl}\partial_it_{jl}
+\frac{g\phi^2}{a^4}\epsilon^{ijk}(t_{kl}\partial_i h_{jl}+h_{kl}\partial_i t_{jl})
\biggr]\,.\nonumber
\end{align}
Note that in the above we have already used the slow-roll approximation $(\dot{\phi}\simeq H\phi)$.
{}From the above second order action one can readily compute $\pi^T_{ij}$%
\be \label{piT}
\pi^T_{~ij}=\left((\frac{\dot{\phi}^2}{a^2}-\frac{g^2\phi^4}{a^4})h_{ij}+2(-\frac{\dot{\phi}}{a}\frac{\dot{t}_{ij}}{a}+
\frac{g^2\phi^3}{a^3}\frac{t_{ij}}{a})+2\frac{g\phi^2}{a^4}\epsilon^{kl}_{~~(i}\partial_k t_{j)l}\right)\,. \ee

Being traceless and divergence-free $t_{ij}$ and $h_{ij}$ each has 2 degrees of freedom which are usually decomposed
into plus and cross ($+$ and $\times$) polarization states  with the polarization tensors $e_{ij}^{+,\times}$ ($\nabla^2 e_{ij}^{+,\times}=-k^2e_{ij}^{+,\times}$).
However, we have parity-violating interaction terms in the action and $\pi^T_{ij}$ which $e_{ij}^{+,\times}$ are not their eigne values. One may then use the right-handed and left-handed circular polarizations and introduce $h_{_{R,L}}$ variables instead%
\bea
h_{ij}=\frac{1}{2a}
  \begin{pmatrix}
   h_{_{R}}+h_{_{L}}& ~-i( h_{_{R}}-h_{_{L}})&~0 \\
   \\
  -i( h_{_{R}}- h_{_{L}})  & ~ -( h_{_{R}}+h_{_{L}})&~0\\
   \\
 ~ 0& ~0&~0
  \end{pmatrix}
\eea
where working with Fourier modes, we chose  $k^i=(0,0,k)$ and imposed the transversality condition.
In a similar way, one can parameterize $t_{ij}$ and $\pi^T_{ij}$ in terms of right and left circular polarizations $T_{_{R,L}}$ and $\pi^T_{_{R,L}}$. Note that, we defined $T_{_{R,L}}$ as the right/ left polarization of the tensor  $\frac{t_{ij}}{a}$.

In terms of $h_{_{R,L}}$ and $T_{_{R,L}}$ and conformal time $\tau$ $(dt=ad\tau$), field equation \eqref{T} reads as
\bea
h_{_{R,L}}''+\big(k^2-(2-\epsilon)\mH^2\big)h_{_{R,L}}\simeq 2a^2\pi^T_{_{R,L}},
\eea
where
\be
a^2\pi^T_{_{R,L}}\simeq2\psi(-\mH T_{_{R,L}}'+\gamma \mH^2 T_{_{R,L}}
\mp k\mH\sqrt{\gamma}T_{_{R,L}})+\mH^2\psi^2(1-\gamma) h_{_{R,L}},
\ee
where prime denotes derivative with respect to the conformal time and $\mH=\dot{a}$.
Moreover, using the second order action, we have the field equation of $T_{_{R,L}}$ as
\be
T''_{_{R,L}}+\bigg(k^2+(2(1+\gamma)+\epsilon)\mH^2\mp2k\mH\frac{1+2\gamma}{\sqrt{\gamma}}\bigg)T_{_{R,L}}\simeq\psi\bigg(\mH h'_{_{R,L}}+\mH^2(1+\gamma) h_{_{R,L}}\mp \sqrt{\gamma}k\mH h_{_{R,L}}\bigg),
\ee

To analyze the tensor modes $h_{_{R,L}}$ and $T_{_{R,L}}$ and the action \eqref{2ndts}, it proves useful to decompose $T_{_{R,L}}$ into $h_{_{R,L}}$ and a new variable $w_{_{R,L}}$%
\be\label{dec-T}
T_{_{R,L}}=A\psi h_{_{R,L}}+w_{_{R,L}}\,,
\ee
where $A$ is a constant (to be determined). In terms of $w_{_{R,L}}$ the equations of motion for $h_{_{R,L}}$ and $T_{_{R,L}}$ read as%
\begin{align}
\label{h-eom}%
h''_{_{R,L}}+(k^2-\frac{z''}{z})h_{_{R,L}}&\simeq4\psi(-\mH w'_{_{R,L}}\mp k\sqrt{\gamma}\mH w_{_{R,L}}+\gamma\mH^2w_{_{R,L}})\cr &+4A\psi^2(-\mH h'_{_{R,L}}+\mH^2 h_{_{R,L}}
\mp k\mH\sqrt{\gamma}h_{_{R,L}})\,,\\
\,\,\cr
\label{w-eom}%
w_{_{R,L}}''+(k^2-\frac{\vartheta''}{\vartheta}\mp 2k\mH \frac{(1+2\gamma)}{\sqrt{\gamma}})w_{_{R,L}}&\simeq
\psi\bigg(\mH h'_{_{R,L}}-\mH^2 h_{_{R,L}}\pm k\mH\frac{~2A(1+2\gamma)-\gamma}{\sqrt{\gamma}}h_{_{R,L}}\nonumber\\
&+(1-2A)(2+\gamma)\mH^2h_{_{R,L}}\bigg)
\end{align}
where%
\bea \frac{z''}{z}&=&\mH^2\bigg(2-\epsilon+2(\gamma-1)\psi^2(2A-1)\bigg)\,,\\
\frac{\vartheta''}{\vartheta}&\simeq&-2\mH^2(\gamma+1)\,. %
\eea%
The above equations imply that  $h_{_{R,L}}$ and $w_{_{R,L}}$ have both oscillatory behavior
$e^{ik\tau}$ in the asymptotic past $k\tau\rightarrow-\infty$ region. However, since $\vartheta''/\vartheta$ is negative while $z''/z$ is positive, they behave differently in superhorizon $k\tau\to 0$ limit; $\frac{h_{_{R,L}}}{a}$ freezes out and $\frac{w_{_{R,L}}}{a}$ decays.
Therefore, in this limit the leading contribution to the right hand side of equation of motion for $w_{_{R,L}}$, which is of order $(k\tau)^{-3}$, should vanish. That is,
\be%
(-2+\frac{1}{A})(2+\gamma)\mH^2 h\simeq 0\,, %
\ee%
which implies $A=\frac{1}{2}$. This choice  for $A$ has an interesting and natural geometric meaning, recalling
the form of our ansatz for the background gauge field, $A^a_{~i}=\psi e^a_{~i}$, where $\psi$ is a scalar (effective inflaton field) and $e^a_i$ are the 3D triads, and that the triads are ``square roots`` of metric.
Perturbing the ansatz and considering only the metric tensor
perturbations $h_{ij}$, we have %
\be
\delta e^a_{~i}=\frac{1}{2}h_{ij}\delta^{aj}\,. %
\ee%
Then, recalling \eqref{gauge-field-pert-2} and the definition of $t_{ij}$, this implies that $A=\frac12$ naturally removes the part of the gauge field tensor perturbations which is coming from the perturbation in the metric, and hence the ``genuine'' gauge field tensor perturbation is parameterized by $w$.

\subsection{Primordial power spectra and the spectral indices}

In the previous part, we  provided the complete set of equations which govern the dynamics of scalar, vector and tensor modes. In this subsection we set about solving these equations, quantize their solutions, and  compute the power spectra.

\subsubsection{Scalar modes}

In order to determine the power spectrum of the scalar perturbations we have to deal with four constraint \eqref{pi-S}, \eqref{dq}, \eqref{drho}, \eqref{A0eq} and one dynamical equation \eqref{dP}. In contrast to the case of scalar field inflationary models which one of the constraints (a combination of $\delta P$ and $\delta q$ equations) reduces to the equation of motion of the background field,
in our case both of them remain independent and should be considered. In Appendix C we provide more details about this issue.

From the combination of \eqref{dq}-\eqref{dP} and \eqref{PiS}, we obtain
\bea\label{Psi-eq}
\ddot\Psi+H\dot\Phi-2\frac{g^2\phi^4}{a^4}\Phi+\frac{k^2}{a^2}(2\mM-\Psi)-2\frac{\dot \phi}{a}(\frac{\dot \mQ}{a}-\frac{\dot\phi}{Ha}\dot\Psi)-4\frac{g^2\phi^3}{a^3}\frac{\mQ}{a}+2\frac{\dot\phi^2}{a^2}\epsilon\Psi=0.
\eea
Furthermore, one can write \eqref{drho} as
\bea\label{drho-eq}
&~&(6H^2-3\YM2)\frac{\dot \mQ}{\dot\phi}+(12H^2-6\frac{\dot\phi^2}{a^2})\frac{\mQ}{\phi}-\frac12\frac{k^2}{a^2}(1+\kk)(\Psi+\Phi)+\frac{k^2}{a^2}(3+\kk)\Psi\nonumber\\
&~&-\frac{k^2}{a^2}(3+\kk+2\dk)\mM+3\YM2\Phi+3\YM2\frac{\Dot\Psi}{H}+3\epsilon\YM2\Psi=0.
\eea
Also using \eqref{PiS}, we can omit $\tM$ in \eqref{dq} and obtain
\be\label{dq-eq}
\dot\Psi+H(1+\frac\gamma2)\Phi-\dot{\mM}+H(1+\gamma)\mM-(\frac{\gamma}{2}-\frac{\phi^2}{a^2})H\Psi-\frac{\dot\phi}{a}\frac{\mQ}{a}\simeq0.
\ee
Equations \eqref{A0eq}, \eqref{Psi-eq}, \eqref{drho-eq} and \eqref{dq-eq} make a complete set of equations for
determining $\mQ, \mM, \Psi$ and $\Phi$. Moreover, using \eqref{pi-S}, one can then determine $\tM$ in terms of the rest of variables.

In order to derive the closed form differential equations governing the dynamics of our dynamical variable $\mQ$, we first write equations in the two asymptotic limits of asymptotic past ($\frac{k}{a}\gg H$) and superhorizon scales ($\frac{k}{a}\ll H$) and then combining them together. Note that, we will rewrite the equations in conformal time $\tau$ ($\tau=\int\frac1a dt$).

\vskip 2mm
\noindent
$\blacktriangleright$ \emph{Asymptotic past limit ($k\tau\gg1$):}

In this limit , constraint equations \eqref{A0eq} and \eqref{dq-eq} take the following forms respectively
\be\label{eq-in-A-P}
k^2(\Psi+\Phi)=0\quad \textmd{and}\quad \mM'-\Psi'+\frac{\dot\phi}{a}\mQ=0,
\ee
here prime denotes a derivative respect to conformal time. From the above equation we find out that the non-zero scalar anisotropic inertia $a^2\pi^S$ is not zero, but given as
\be\label{piS-A-P}
a^2\pi^S=-2\Phi.
\ee
We note that regardless of the details, for all the scalar inflationary models in the context of GR the anisotropic stress
$a^2\pi^S$ is identically zero. Thus, the non-zero anisotropic inertia in our system is directly related to the existence of
gauge fields in the set up.

Using the constraints in \eqref{energy-conservation} and \eqref{Psi-eq}, we then can omit $\Psi$ and $\mM$ in terms of $\mQ$ and $\Phi$ which leads to the following set of coupled equations for $\Phi$ and $\mQ$ respectively
\bea\label{Q-Psi-1}
&~&\frac{\dot\phi}{a}\mQ''+k^2\big(\frac{\gamma+2}{3\gamma}\frac{\dot\phi}{a}\mQ+\frac{2}{3\gamma}\Phi'\big)=0,\\\label{Q-Psi-2}
&~&(2\frac{\dot\phi}{a}\mQ+\Phi')''+k^2\big(2\frac{\dot\phi}{a}\mQ+\Phi'\big)=0.
\eea
These equations imply that in the asymptotic past limit, we have $\mQ\propto k\Phi$.
As we see, the first equation has a complicated form, however the second one is simply a wave equation for $~2\frac{\dot\phi}{a}\mQ+\Phi'~$ with a sound speed equal to one.
Multiplying the former by a factor of $(\gamma+1)$ and subtracting the result from the latter, we obtain the following wave equation for the variable $~(\gamma-1)\frac{\dot\phi}{a}\mQ-\Phi'~$
\be
(\gamma-1)\frac{\dot\phi}{a}\mQ''-(\Phi')''+(\frac{\gamma-2}{3\gamma})k^2\big((\gamma-1)\frac{\dot\phi}{a}\mQ-\Phi'\big)=0,
\ee
with a sound speed square equal to $(\frac{\gamma-2}{3\gamma})$.
In other words, in this limit\footnote{Defining $X_1=2\frac{\dot\phi}{a}\mQ+\Phi'\quad \textmd{and} \quad X_2=(\gamma-1)\frac{\dot\phi}{a}\mQ-\Phi'$,
we can diagonalize the set of equations \eqref{Q-Psi-1}-\eqref{Q-Psi-2} into two wave equations for
 for $X_1$ and $X_2$ with sound speeds $c_1^2=1$ and $c_2^2=(\frac{\gamma-2}{3\gamma})$ respectively.}
\begin{itemize}
\item{One can decompose $\mQ$ as $\mQ=\mQ_1+\mQ_2$ where $\mQ_{1,2}$ satisfy
\bea
\label{Q+-}
\mQ_{1}''+k^2\mQ_{1}= 0\,,\quad  \quad \mQ_{2}''+\frac{\gamma-2}{3\gamma}k^2\mQ_{2}= 0.
\eea
}
\item{Then, we can decompose $\Phi$ as $\Phi=\Phi_1+\Phi_2$ such that
\bea
\label{Psi+-}
\Phi_{1}''+k^2\Phi_{1}= 0\,,\quad  \quad \Phi_{2}''+\frac{\gamma-2}{3\gamma}k^2\Phi_{2}= 0.
\eea
Besides that, we also have the following two constraints
\be
\Phi'_1=(\gamma-1)H\psi \mQ_1\,,\quad \quad  \Phi'_2=-2H\psi \mQ_2,
\ee
which in the asymptotic past limit, couple $\Phi$ and $\mQ$ fields. }
 \end{itemize}

\vskip 2mm
\noindent
$\blacktriangleright$ \emph{The superhorizon limit ($k\tau\to 0$)}

We now turn to the question of large scale superhorizon behavior of the system in $k\tau\to 0$.
In the superhorizon limit, \eqref{A0eq} and \eqref{drho-eq} take the following forms respectively
\bea
\label{A0eq-S-H}
&&\frac{\mQ}{\phi}+\frac{\delta}{\epsilon}\times(\Phi+\frac{\dot\Psi}{H})-\frac12\frac{g^2\phi^4}{a^4H^2}\Psi=0,\\
\label{drho-eq-S-H}
&&\frac{\mQ}{\phi}+\frac16(\epsilon-\eta)(\Phi+\frac{\dot\Psi}{H})\simeq0.
\eea
Combining of the above equations and using the slow-roll background relation \eqref{eta-x}, we obtain
\be
\label{super-cons}
\frac{\mQ}{\phi}\simeq-\frac16(\epsilon-\eta)\Phi\quad\textmd{and}\quad \frac{\mQ}{\phi}\sim\Psi,
\ee
which indicates that at the superhorizon limit, the scalar anisotropic stress
$a^2\pi^S$ is non-vanishing and is given by
\be
a^2\pi^S\simeq-\Phi.
\ee
From \eqref{piS-A-P}, we found that gauge-flation has a non-zero $a^2\pi^S$ at the asymptotic past limit.
Now, the above relation indicates that this quantity has a non-zero value also at the superhorizon which makes it an observable quantity. This is a unique and specific feature of the non-Abelain gauge field
inflation, not shared by any scalar-driven inflationary model.

Use of the above result in \eqref{Psi-eq}, we have the following equation for $\Phi$
\be
\label{super-Phi-eq}
\Phi''-2(\epsilon-\eta)\mH^2\Phi\simeq0,
\ee
while \eqref{energy-conservation} leads to the following equation for $\mQ$
\bea
\label{super-Q-eq}
\mQ''-\mH^2(2+8\epsilon+6(\epsilon-\eta))\mQ\simeq 0.
\eea

As we see, in this limit, we have only one equation for both of $\mQ_1$ and $\mQ_2$,
similarly both of $\Phi_1$ and $\Phi_2$ are described by the same equation.
Besides, similar to all the other adiabatic perturbations, the Bardeen potentials $\Phi$ and $\Psi$ are both constant on super-Hubble scales $(k\tau\ll1)$.

{}Up to now we worked out the field equations of $\mQ$ and $\Phi$ in asymptotic past and the superhorizon limits. In the following we combine them and read the closed form differential equations corresponding to each field and study the system.

Upon using \eqref{Psi+-}, \eqref{super-Phi-eq}, $\Phi=\Phi_1+\Phi_2$ is governed by the following two dynamical equations
\bea
\label{Phi+}
\Phi_{1}''+\big(k^2-\frac{\theta''}{\theta}\big)\Phi_{1}&\simeq &0,\\
\label{Phi-}
\Phi_{2}''+\big(\frac{\gamma-2}{3\gamma}k^2-\frac{\theta''}{\theta}\big)\Phi_{2}&\simeq &0,
\eea
which their solutions subject to the following constraint equation
\be\label{constraint-Phi-eq}
\Phi'_1\simeq(\gamma-1)\psi HQ_1,\quad \textmd{at}\quad  \Phi'_2\simeq-2\psi HQ_2\,,
\ee
in the asymptotic past limit. Here $\frac{\theta''}{\theta}=2\mH^2(\epsilon-\eta)$, which can be written as
\be\label{nu-Phi}
\frac{\theta''}{\theta}=\frac{\nu^2_{R}-\frac14}{\tau^2},\quad \textmd{where} \quad \nu_{R}\simeq\frac12+2(\epsilon-\eta).
\ee
Note that in determining the above relation, we used the slow-roll approximation $$\tau\simeq-\frac{1}{(1-\epsilon)H}.$$

The general solutions to \eqref{Phi+} and \eqref{Phi-} for $\Phi_{1,2}$ can be expressed as a linear combination
of Hankel $H^{(1)}_\nu$ and $H^{(2)}_\nu$, and modified Bessel functions $I_\nu$ and $K_\nu$. Recalling \eqref{nu-Phi}, this leads to the following solutions for $\Psi$
\be
\Phi_{1}(k,\tau)\simeq \frac{\sqrt{\pi\vert\tau\vert}}{2k}\big(b_1 H^{(1)}_{\nu_R}(k\vert\tau\vert)+\tilde{b}_1 H^{(2)}_{\nu_R}(k\vert\tau\vert)\big),
\ee
and
\bea\label{Phi2}
\Phi_{2}(k,\tau)\simeq \left\{\begin{array}{cc}\frac{\sqrt{\vert\tau\vert}}{\sqrt{\pi}k}\big(b_2 K_{\nu_R}(\sqrt{\frac{\vert 2-\gamma\vert}{3\gamma}}k\vert\tau\vert)+\tilde{b}_2 I_{\nu_R}(\sqrt{\frac{\vert 2-\gamma\vert}{3\gamma}}k\vert\tau\vert)\big), &\quad \gamma-2<0\\
\frac{\sqrt{\pi\vert\tau\vert}}{2k}\big(ib_2 H^{(1)}_{\nu_R}(\sqrt{\frac{\vert \gamma-2\vert}{3\gamma}}k\vert\tau\vert)+\tilde{b}_2 H^{(2)}_{\nu_R}(\sqrt{\frac{\vert\gamma-2\vert}{3\gamma}}k\vert\tau\vert)\big),&\quad \gamma-2>0.
\end{array}\right.
\eea
In \eqref{Phi2}, the coefficients are chosen in such a way that for both cases $\Phi_2$ satisfies \eqref{constraint-Phi-eq} with the same value.

\vskip 2mm
\noindent
$\blacktriangleright$ \emph{\textbf{Classical solutions for $\mQ$.}}

The second order action computation is indeed very tedious, lengthy and cumbersome, but that is necessary for quantization of the perturbations. This is because for performing the canonical quantization of the modes, besides the equations of motion we need to have the canonical (conjugate) momentum too. In the Appendix A we have presented the explicit form of the second-order action, after imposing the gauge-fixing conditions ($E=B=0$).

Appearance of negative $c^2_s$ modes may cause a concern about a possibility of ghost instability in our system.
Theoretically we do not expect finding ghosts in our theory because, \emph{i)} we are dealing with a gauge-invariant action and we respect this gauge symmetry. (To be more precise, it is spontaneously broken by the choice of classical inflationary background. However, as is well-established, spontaneous gauge symmetry breaking does not lead to a break-down of Slavnov-Taylor identity which reflects the gauge symmetry and its consequences about renormalizability and unitarity.) \emph{ii)} Although we are dealing with a ``higher derivative'' action \eqref{The-model}, the higher derivative term has a special form: it does not involve more than time-derivative squared terms. (This fact is also explicitly seen in \eqref{phi-eom} in that the $\phi$ equation of motion  does not involve more than second time derivative.) As such we expect not to see ghosts usually present in the higher derivative theories.
Besides the above arguments, to make sure about the absence of ghosts, we have explicitly computed the second-order action. The expression for the second-order action, after implementing the constraints, explicitly shows that neither $\mQ$ nor $\mM$ has negative kinetic terms and hence there is no ghost instability in our system.
The explicit expression for the second-order action is presented in Appendix A and here we only present the simplified result in the asymptotic past limit.

After combining \eqref{Q+-} and \eqref{super-Q-eq}, the $\mQ$ equations of motion in the slow-roll approximation
 take the form
\bea
\label{Q+}
&&\mQ_{1}''+\big(k^2-\frac{z''}{z}\big)\mQ_{1}\simeq0,\\
\label{Q-}
&&\mQ_{2}''+\big(\frac{\gamma-2}{3\gamma}k^2-\frac{z''}{z}\big)\mQ_{2}\simeq0,
\eea
where the effective mass term is given as $$\frac{z''}{z}\simeq (2+8\epsilon+6(\epsilon-\eta))\mH^2.$$
Moreover, the solutions subject to the following algebraic constraint at superhorizon scales \eqref{super-cons}
\be\label{algebraic-Q-eq}
\mQ_{1}+\mQ_{2}=\mathcal{O}(\epsilon)\phi\Phi\quad \textmd{at}\quad  k\tau\ll1\,.
\ee
On the other hand, up to the leading orders in slow-roll ($\mH\simeq-(1+\epsilon)/\tau$),
we can write $\frac{z''}{z}$ as
\be\label{nu-R}
\frac{z''}{z}=\frac{\nu_Q^2-\frac14}{\tau^2}\quad \textmd{where}\quad \nu_Q\simeq\frac32+2(3\epsilon-\eta).
\ee
The general solution to the equation \eqref{Q+} is a linear combination of Hankel functions
\be
\mQ_{1}(k,\tau)\simeq \frac{\sqrt{\pi\vert\tau\vert}}{2}e^{i(1+2\nu_Q)\pi/4}\big(q_1 H^{(1)}_{\nu_Q}(k\vert\tau\vert)+\tilde{q}_1 H^{(2)}_{\nu_Q}(k\vert\tau\vert)\big).
\ee
On the other hand, the general solution of \eqref{Q-} is
\bea\label{Q2}
\mQ_{2}(k,\tau)\simeq \left\{\begin{array}{cc}\frac{\sqrt{\vert\tau\vert}}{\sqrt{\pi}}\big(q_2 K_{\nu_Q}(\sqrt{\frac{\vert 2-\gamma\vert}{3\gamma}}k\vert\tau\vert)+\tilde{q}_2 I_{\nu_Q}(\sqrt{\frac{\vert 2-\gamma\vert}{3\gamma}}k\vert\tau\vert)\big), &\quad \gamma-2<0\\
\frac{\sqrt{\pi\vert\tau\vert}}{2}\big(iq_2 H^{(1)}_{\nu_Q}(\sqrt{\frac{\vert\gamma-2\vert}{3\gamma}}k\vert\tau\vert)+\tilde{q}_2 H^{(2)}_{\nu_Q}(\sqrt{\frac{\vert\gamma-2\vert}{3\gamma}}k\vert\tau\vert)\big),&\quad \gamma-2>0,
\end{array}\right.
\eea
which as we see in case that $\gamma-2<0$, it is expressed as a linear combination of modified Bessel functions,
otherwise it is expressed in terms of Hankel functions. Note that in \eqref{Q2}, the coefficients are chosen such
that in both cases, $\mQ_{2}$ has the same superhorizon value.

\vskip 2mm
\noindent
$\blacktriangleright$ \emph{\textbf{Quantization of  $\mQ$ modes.}}

As in the standard text book material in cosmic perturbation theory, the coefficients $b_i, \tilde{b}_i, q_i$ and $\tilde{b}_i$ may be fixed using the
canonical normalization of the modes in the Minkowski, deep subhorizon $k\tau\to -\infty$ regime. As discussed, in this limit  $\mQ_1$, which has an oscillatory behavior, is the only quantum field. We should stress that, of course not all coefficients are fixed by the quantization normalization condition. To fix them, as we will do so below, we should impose the constraints \eqref{constraint-Phi-eq} and \eqref{algebraic-Q-eq} in both superhorizon and asymptotic past regimes. Note also that fulfilling these constraints is equivalent to maintaining the diffeomorphism and remainder of the gauge symmetry of the system; fluctuations both at classical and quantum levels must respect them.

{}From the second-order action given in Appendix A, after using the constraints and some lengthy straightforward algebra, we determine the form of the 2nd order action at the asymptotic past limit
 \bea\label{2ndS-asymtotic-past}%
 \delta_2S_{_{tot}}&\simeq&\int
d\tau d^3x
 \Biggl[(1+\gamma)(\mQ_1'^2-k^2\mQ_1^2)+3\frac{(\gamma+1)}{(\gamma-2)}(\mQ_2'^2-\frac{(\gamma-2)}{3\gamma}k^2\mQ_2^2)\Biggl].
 \eea%
Now we can read the canonically normalized field value which is given as
 \be\label{Canonically-norm}
 \mQ_{_{norm}}=\sqrt{2(1+\gamma)}\mQ_1.
 \ee
Imposing the usual Minkowski vacuum state for
$\mQ_{_{norm}}$%
$$
\mQ_{_{norm}}\simeq \frac{1}{\sqrt{2k}}e^{-ik\tau}\,,
$$
fixes the $q_1$, $\tilde{q}_1$ and $\tilde{q}_2$  coefficients
\be\label{q1}
q_1 =\frac{1}{\sqrt{2(\gamma+1)}}\,,\qquad \tilde{q}_1=\tilde{q}_2= 0\,.
\ee
It is useful to remind that the $H^{(1)}_\nu(z)$ and $K_\nu(z)$ functions have the following asymptotic forms in the limit of $z\gg 1$:$$H^{(1)}_\nu(z)\simeq\sqrt{\frac{2}{z\pi}}e^{-\frac{i\pi}{4}(2\nu+1)} e^{iz},\quad K_\nu(z)\simeq\sqrt{\frac{2}{z\pi}}e^{-z}.$$
Moreover, their asymptotic forms in the limit of $z\ll 1$ is as follows: $$H^{(1)}_\nu(z)\simeq-\frac{i}{\pi}\Gamma(\nu)\big(\frac{z}{2}\big)^{-\nu},\quad K_\nu(z)\simeq\frac{1}{2}\Gamma(\nu)\big(\frac{z}{2}\big)^{-\nu}.$$
From \eqref{algebraic-Q-eq} and after using the asymptotic forms of $H^{(1)}_\nu(z)$ and $K_\nu(z)$ in the $z\ll 1$ limit, one can read $q_2$ as
\be\label{q2}
q_2\simeq i\bigg(\frac{\vert\gamma-2\vert}{3\gamma}\bigg)^\frac{3}{4}q_1=\frac{i}{\sqrt{2(1+\gamma)}}\bigg(\frac{\vert\gamma-2\vert}{3\gamma}\bigg)^\frac{3}{4}.
\ee
Obtaining $q_{1,2}$ and $\tilde{q}_{1,2}$, now we turn to determine the Bardeen potential $\Phi$.

Putting \eqref{q1} and \eqref{q2} into \eqref{constraint-Phi-eq} and after using the asymptotic form of Bessel functions in the $k\tau\rightarrow-\infty$, we have $\tilde{b}_1=\tilde{b}_2=0$, while
\be
b_1\simeq-\frac{(\gamma-1)}{\sqrt{2(\gamma+1)}}H\psi,\quad \textmd{and}\quad b_2\simeq \frac{2i}{\sqrt{2(1+\gamma)}}\bigg(\frac{\vert\gamma-2\vert}{3\gamma}\bigg)^{\frac14}H\psi.
\ee
Now we are ready to determine the superhorizon value of the Bardeen potential $\Phi$.
Having the coefficients above and using the background slow-roll relation $\epsilon\simeq(1+\gamma)\psi^2$,  we obtain
\be\label{after-horizon-Phi}
\Phi=\Phi_1+\Phi_2\simeq\frac{i\sqrt{\epsilon}}{2k^{3/2}}H\big(\frac{k\vert\tau\vert}{2}\big)^{\frac12-\nu_R},\quad k\vert\tau\vert\ll1.
\ee
In the asymptotic past limit $\Phi_2$
is a mode which can have negative $c_s^2$ for $\gamma<2$. Nonetheless, our analysis above shows explicitly that this
does not render our perturbation theory analysis unstable, because what is physical is the total $\Phi$ \emph{after} imposing
the constraint equations on the \emph{superhorizon}
scales. In other words, $\Phi_2$ mode in the asymptotic past is fixed by the constraints on the dynamical equations
and not an independent mode.

\vskip 2mm
\noindent
$\blacktriangleright$ \emph{\textbf{The curvature power spectrum.}}

Having fixed all the coefficients, we are now ready to compute the power spectrum of metric and curvature perturbations. The power spectrum for the metric perturbations is given by \cite{Inflation-Books} %
\be
\mathcal{P}_\Phi=\frac{4\pi k^3}{(2\pi)^3}|\Phi|^2\,,
\ee
which on  large superhorizon scales ($k\ll aH$) is %
\be
\mathcal{P}_\Phi\simeq\frac{\epsilon}{8}\left(\frac{H}{\pi}\right)^2
\left(\frac{|k\tau|}{2}\right)^{3-2\nu_\mQ}\,,
\ee
and  remains constant during slow-roll period. The power spectrum of the comoving curvature perturbation ${\cal R}$,
$\mathcal{R}\simeq\frac{\Phi}{\epsilon}$, is hence%
\be\label{PR}
\mathcal{P}_{\mathcal{R}}\simeq\frac{1}{8\epsilon}\left(\frac{H}{\pi}\right)^2|_{k=aH}\,, \ee%
and becomes constant on super-Hubble scales.
 Note that the scalar power spectrum in our model is exactly equal to the power spectrum of the
comoving curvature perturbation in the standard single scalar field model.

The spectral index of the curvature perturbations, $n_{\mathcal{R}}-1=3-2\nu_{\mathcal{Q}}$, to the
leading order in the slow-roll parameters is%
\be\label{nr}
n_{\mathcal{R}}-1\simeq -2(\epsilon-\eta)\,,
\ee
We note that the spectral tilt \eqref{nr} is always negative in our model.

In addition to the power spectrum of the scalar and its spectral tilt, our model has a non-zero scalar anisotropic stress value with the following power spectrum
\be\label{PpiS}
\Delta_{a\!^2\!\pi^S}^2\simeq\frac{\epsilon}{8}\left(\frac{H}{\pi}\right)^2\!\!\bigg{|}_{k=aH}\,,
\ee%
which becomes constant on super-Hubble scales. Thus, as one of the specific features of the
non-Abelian gauge field inflation, power spectrum of scalar anisotropic stress is non-zero.
This is in contrast with all  scalar-driven inflationary models in the general relativity,
for which $a^2\pi^S$ is identically zero.

\subsubsection{Tensor modes}

\begin{figure}[t]
\begin{center}
\includegraphics[angle=0, width=80mm, height=70mm]{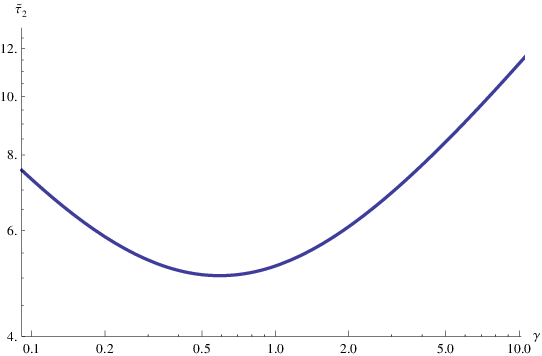}
\caption{$\tilde{h}_{R}$ undergoes a tachyonic growth phase in $\tilde{\tau}_2(\gamma)\geq -k\tau \geq 1$ (\emph{cf}.  \eqref{tilde-h-simple} and \eqref{tilde-h-omega}). In this figure, we have depicted $\tilde{\tau}_2$ vs. $\gamma$. The minimum is $\tilde\tau=5$ which is at $\gamma\simeq0.6$.}\label{omega-root-fig}
\end{center}
\end{figure}

\begin{figure}[ht]
\includegraphics[angle=0, width=80mm, height=78mm]{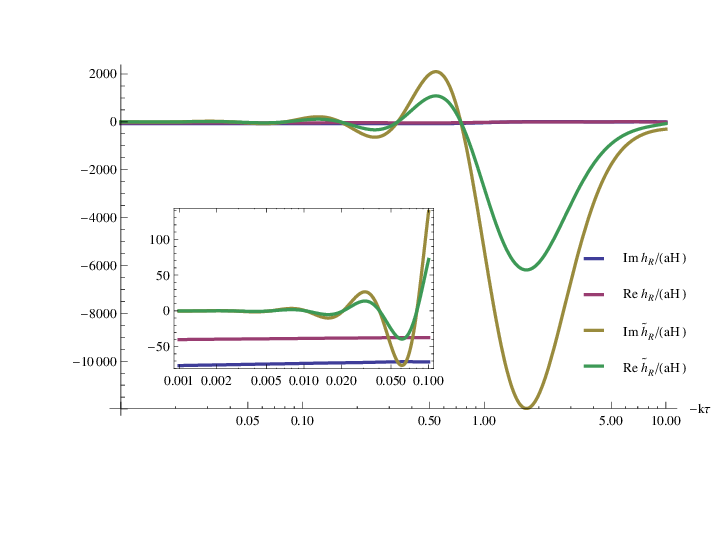}
\includegraphics[angle=0, width=80mm, height=78mm]{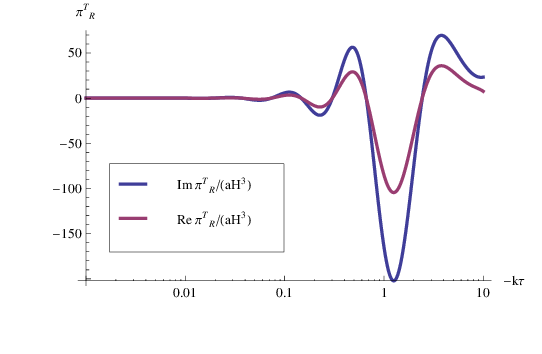}\\
\includegraphics[angle=0,width=80mm, height=78mm]{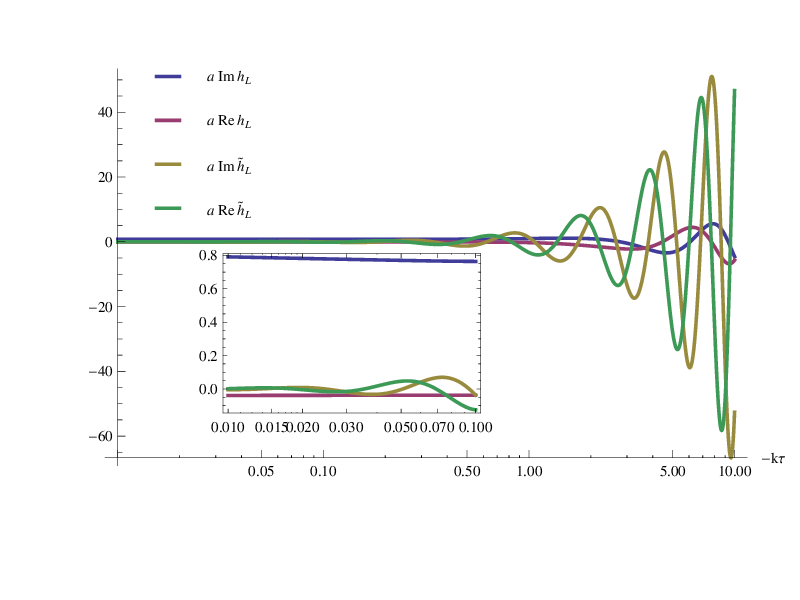}
\includegraphics[angle=0,width=80mm, height=78mm]{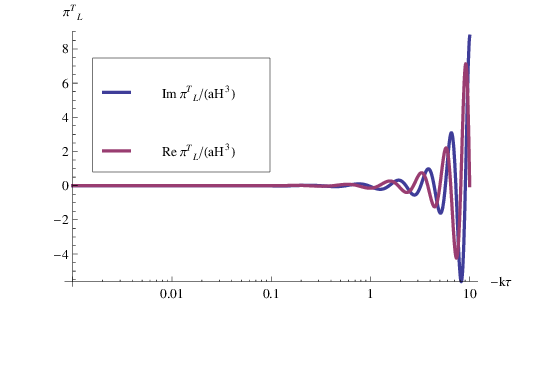}
\caption{This figure presents the tensor modes solution for $\psi=5\times 10^{-2}$, $\gamma=10$ and $H_0=10^{-6}$. In the top-left panel, we have the tensor field values $\frac{h_{_{R}}}{aH}$ and $\frac{\tilde h_{_{R}}}{aH}$ versus $-k\tau$, where \textit{Re} and \textit{Im} denote read and imaginary parts of the corresponding quantity. The small box presented the superhorizon behavior of the fields. The top-right panel shows $\frac{\pi^T_{_{R}}}{aH^3}$. In the bottom panels we presented the left-handed polarizations.}\label{Tensor-Mode-figures}
\end{figure}

In the previous section we found that upon setting $A=\frac{1}{2}$ in \eqref{dec-T} field equations of $h_{_{R,L}}$ and $w_{_{R,L}}$ decouple at the superhorizon scales. Then, the  equation of $h_{_{R,L}}$ fields read as
\bea\label{h}
h_{_{R,L}}''+\big(k^2-(2-\epsilon)\mH^2\big)h_{_{R,L}}\simeq 2a^2\pi^T_{_{R,L}},
\eea
where
\be
a^2\pi^T_{_{R,L}}\simeq\psi(-2\mH w_{_{R,L}}'+2\gamma \mH^2 w_{_{R,L}}
\mp2k\mH\sqrt{\gamma}(w_{_{R,L}}+\frac12\psi h_{_{R,L}})-\mH \psi h_{_{R,L}}'+\mH^2\psi h_{_{R,L}}).
\ee
The above equations imply that in the superhorizon limit, we have $h_{_{R,L}}\propto a$.
On the other hand,  field equations of $w_{_{R,L}}$ leads to
\be\label{eq-tildeh}
w_{_{R,L}}''+\left(k^2+\big(2(1+\gamma)+\epsilon\big)\mH^2\mp2k\mH\frac{(1+2\gamma)}{\sqrt{\gamma}}\right)w_{_{R,L}}\simeq
 \mH\psi\left[(h_{_{R,L}}'-\mH h_{_{R,L}}) \pm k\frac{(\gamma+1)}{\sqrt{\gamma}}h_{_{R,L}}\right]\,,
\ee
which implies that while $w_{_{R,L}}$ behaves like a plane-wave at subhorizon scales,
it is exponentially damped like $w_{_{R,L}}\propto a^{-(1+\gamma)}$ at superhorizon scales ($h_{_{R,L}}\propto a$). Although a damping mode at superhorizon scales, $w_{_{R}}$ has an interesting behavior just before the horizon crossing. In fact, the parity violating terms in the field equation of $w_{_{R}}$ leads to tachyonic growth of $w_{R}$ around the horizon crossing. To see this, let us neglect $h_{_{R,L}}$ terms 
in the RHS of \eqref{eq-tildeh} to find the following wave equation for $w_{_{R,L}}$
\be\label{tilde-h-simple}
\partial^2_{\tilde{\tau}}w_{_{R,L}}+\Omega^2_{_{R,L}}(\tilde{\tau},\gamma)w_{_{R,L}}\simeq0,
\ee
here $\tilde{\tau}=-k\tau$ and $\Omega^2_{_{R,L}}(\tilde{\tau},\gamma)$ is given as
\be\label{tilde-h-omega}
\Omega^2_{_{R,L}}(\tilde{\tau},\gamma)=\bigg(1+\frac{2(1+\gamma)}{\tilde{\tau}^2}\mp2\frac{(1+2\gamma)}{\sqrt{\gamma}\tilde{\tau}}\bigg).
\ee
Note that while $\Omega^2_{_{L}}$ is always positive, $\Omega^2_{_{R}}$ becomes negative  in an interval $\tilde{\tau}\in(\tilde\tau_1,\tilde\tau_2)$. Fig. \ref{omega-root-fig} presents  $\tilde\tau_2$ vs. $\gamma$ and $\tilde\tau_1$ is almost one.
The short interval of negative $\Omega^2_{_{R}}$, leads to the tachyonic growth of $w_{_{R}}$ (Fig. \ref{Tensor-Mode-figures}).\footnote{We had missed this behaviour in the earlier version of this work. A similar feature was pointed out in the context of chromo-natural inflation model in \cite{AMW}. }

From \eqref{h}, we learn that the anisotropic inertia $\pi^T_{_{R,L}}$ is the source term for $h_{_{R,L}}$, which vanishes at superhorizon scales ($k\tau\rightarrow0$). Nonetheless, due to the tachyonic growth of $w_R$ just before the horizon-crossing, $\pi^T_{_{R}}$  has the behavior of an impulse function in that region (see Fig. \ref{Tensor-Mode-figures}), inducing the growth in $h_{_{R}}$ and enhancing its superhorizon value. On the other hand, $\pi^T_{_{L}}$ is small at the horizon crossing and has negligible effect on the superhorizon value of $h_{_{L}}$.

\begin{figure}[t]
\includegraphics[angle=0, width=80mm, height=80mm]{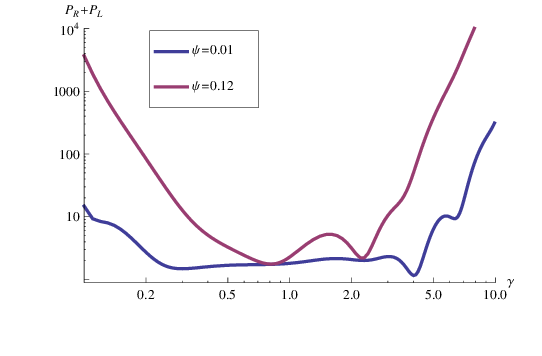}
\includegraphics[angle=0, width=80mm, height=80mm]{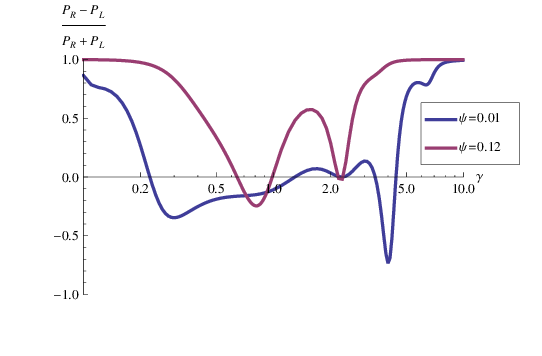}\\
\caption{In the left panel we have depicted ${P_{_{R}}}+{P_{_{L}}}$. In the standard scalar-driven inflationary models ${P_{_{R}}}={P_{_{L}}}=1$. The right panel the parity violating factor $\frac{P_{_{R}}-P_{_{L}}}{P_{_{R}}+P_{_{L}}}$ versus $\gamma$ for $\psi=10^{-2}$ and $\psi=0.12$ is shown. The power spectra have been calculated at $k\tau=-0.01$, long enough after modes have crossed the horizon and behave quite classically. As we see in the right panel,  for very small and very large $\gamma$ values $P_{_{R}}\gg P_{_{L}}$.}\label{Tensor-Mode-power-figures}
\end{figure}

Considering the standard Minkowski (Bunch-Davis)
vacuum normalization for the canonically normalized fields (\emph{cf}. \eqref{2ndts}), leads to
\be
h_{_{R,L}}\rightarrow\frac{e^{-ik\tau}}{\sqrt{k}}\quad \textmd{and} \quad w_{_{R,L}}\rightarrow\frac{e^{-ik\tau}}{2\sqrt{k}}
\,,\qquad k\tau \to -\infty\,.
\ee
The power spectra for the Left and Right gravitational wave modes are obtained as
\bea\label{PT}%
\mathcal{P}_{T_{R}}\simeq  P_{_R} \left(\frac{H}{\pi}\right)^2\!\!\big\vert_{k=aH}\quad\textmd{and}\quad
\mathcal{P}_{T_{L}}\simeq  P_{_L} \left(\frac{H}{\pi}\right)^2\!\!\big\vert_{k=aH}\,,%
\eea%
where $P_{_R},\ P_{_L}$ are  functions of the parameters $\gamma, \psi$ and the power spectrum of the tensor modes is given as
$$\mathcal{P}_{T}=\mathcal{P}_{T_{R}}+\mathcal{P}_{T_{L}}=(P_{_R}+P_{_L})\left(\frac{H}{\pi}\right)^2\!\!\big\vert_{k=aH}.$$
In the left panel of Fig. \ref{Tensor-Mode-power-figures}, we presented $P_{_{R}}+P_{_{L}}$ vs. $\gamma$ and in the right panel of this figure, we have the parity violating ratio $\frac{P_{_{R}}-P_{_{L}}}{P_{_{R}}+P_{_{L}}}$ vs. $\gamma$.
Studying the system numerically, we find that $P_{_L}$ is a function very close to one (ranging from $1.0$ at low $\gamma$ to $1.25$ at $\gamma=10$) while $P_{_R}$ varies significantly in this range of $\gamma$.

Moreover, the spectral index of tensor perturbations, $n_T$ is given by %
\be %
n_T\simeq -2\epsilon\,, %
\ee%
which is equal to its corresponding quantity in the standard scalar inflationary models.
Note that due its exponential suppression on the superhorizon scales, the $w_{_{R,L}}$ mode does not contribute to
the tensor power spectrum.
For our model tensor-to-scalar ratio $r$ is%
\be\label{ratio}%
r=8(P_{_R}+{P_{_L}})\epsilon\,,%
\ee
which can be written as $r=-4(P_{_R}+{P_{_L}})n_T$. That is, our model respects a modified version of the Lyth consistency relation \cite{Lyth-bound}.

\section{Fitting Gauge-flation results with the cosmic data}\label{testing-the-model}

We are now ready to confront our model with the observational data. As discussed our model allows for slow-roll inflation for a specific range of its parameters  and for comparison with the observational data we use the results obtained in the slow-roll regime.
First, we note that in order for inflation to solve the flatness and horizon problems it should have lasted for a minimum number of e-folds $N_e$. This amount of course depends on the scale of inflation and somewhat on the details of physics after inflation ends \cite{Inflation-Books}. However, for a large inflationary scale, like $H\sim 10^{-4}-10^{-5}\mpl$, it is usually demanded that $N_e\simeq 60$. As a standard benchmark we use $N_e\geq50$.

As for the CMB data, current observations provide values for power spectrum of curvature perturbations $\mathcal{P}_{{\cal R}}$ and its spectral tilt $n_{\cal R}$ and impose an upper bound on the power spectrum of tensor modes $\mathcal{P}_{T}$, or equivalently an upper bound on tensor-to-scalar ratio $r$. These values vary (mildly) depending on the details of how the data analysis has been carried out.  Here we use
the best estimation of Komatsu et al. \cite{CMB-data} which is based on WMAP 7 years, combined with other cosmological data. These values are
\begin{subequations}\label{CMB-data}%
\begin{align}
\mathcal{P}_{{\cal R}}&\simeq 2.5\times 10^{-9}\,,\\
n_{\cal R}&= 0.968\pm0.012\,,\\
r&<0.24\,.
\end{align}
\end{subequations}

Our model has two parameters $g$ and $\kappa$, and our results for physical observable depend also on others parameters which are basically related to the initial values of the fields we have in our model. Out of these parameters we choose $H$, the value of Hubble, and $\psi$, the value of the effective inflaton field  at the beginning of, or during, slow-roll inflation. The values of other parameters, $\epsilon,\ \gamma$ and $\delta$ (initial velocity of the $\psi$ field \eqref{delta-def}), are related to these two through \eqref{epsilon-x}-\eqref{x-def}. For convenience let us recollect  our results:
\bea
\label{Ne-section6}N_e&=&\frac{\gamma+1}{2\epsilon}\ln \frac{\gamma+1}{\gamma}\,,\\
\label{nR-section6}n_{\mathcal{R}}-1&\simeq& -\frac{\gamma}{4(\gamma+1)}\frac{r}{(P_{_{R}}+P_{_{L}})}\,,\\
\label{r-section6}r&=&\frac{\mathcal{P}_{T}}{\mathcal{P}_{{\cal R}}}=8(P_{_{R}}+P_{_{L}})\epsilon\,,\\ %
\label{PR-section6}\mathcal{P}_{\mathcal{R}}&\simeq &\frac{1}{8\pi^2\epsilon}\left(\frac{H}{\mpl}\right)^2 \simeq
\frac{g^2}{8\pi^2\gamma(\gamma+1)}\,.
\eea%

\begin{figure}
\begin{center}
\includegraphics[angle=0, width=90mm, height=65mm]{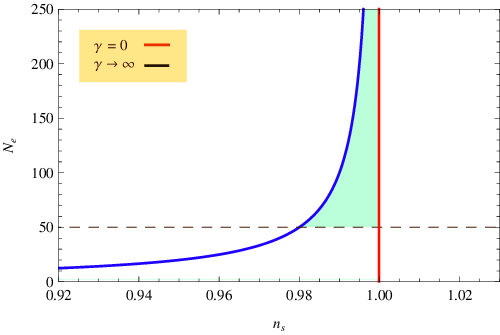}
\caption{The shaded region exhibits the region which leads to $N_e>50$. As we see this may happen for any value of $\gamma$ parameter. This also shows that our spectral tilt is always in $0.98\leq n_s< 1$ range. Moreover, our model allows for arbitrary large $N_e$.}\label{fig-ns-vs-Ne}
\end{center}
\end{figure}

Moreover, from the combination of \eqref{nr} and \eqref{Ne-section6}, one can read $n_s$ in terms of $N_e$ and $\gamma$ as below
\be
n_s\simeq1-\frac{\gamma}{N_e}\ln(\frac{\gamma+1}{\gamma}).
\ee
Since, $\gamma\ln(\frac{\gamma+1}{\gamma})$ is a quantity between zero and one, in gauge-flation model spectrum cannot be very red. In Fig. \ref{fig-ns-vs-Ne}, we presented $N_e$ vs. $n_s$ which indicates that for  $N_e\geq 50$ leads to
\be
n_s^{(50)}\geq 0.98\,.
\ee

\begin{figure}[ht]
\includegraphics[angle=0, width=90mm, height=80mm]{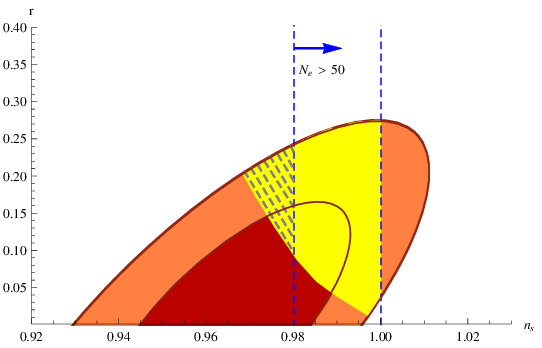}
\includegraphics[angle=0, width=70mm, height=80mm]{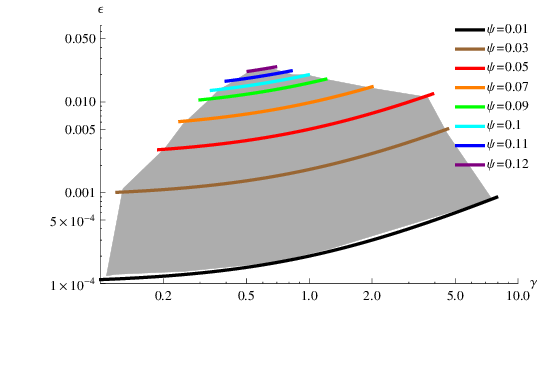}
\caption{The left panel shows  $1\sigma$ and $2\sigma$ contour bounds of 7-year WMAP+BAO+H0. The yellow area (region with lighter color) represents the gauge-flation predictions for $\psi\in(0.01,0.12)$ range. As depicted in Fig. \ref{fig-ns-vs-Ne}, the region with enough number of e-folds restricts us to $n_s>0.98$ region, that is on the right-side of the $N_e=50$ line. Therefore, the allowed region is the highlighted region between $N_e=50$ and $n_s=1$ lines. The shaded region in right panel shows the allowed values for $\epsilon$ and $\psi$, given in \eqref{psi-range} and \eqref{bound-epsilon-gamma}. }\label{ns-r-gf}
\end{figure}

Similarly, as depicted in the left panel of Fig. \ref{ns-r-gf}, we see that  our model predicts a minimum value for $r$
\be\label{r-bound-gf}
0.02\leq r\leq 0.28\,,
\ee
which is a very specific prediction of our model and gauge-flation may be falsified by the upcoming Planck satellite results.
In the allowed region, we have
\be\label{psi-range}
\psi\simeq (0.01-0.1)\mpl.
\ee
 The \textit{max}   and \textit{min} possible values of $r$  respectively correspond to $(\psi=0.01, \gamma=5, P_{_{R}}+P_{_{L}}=6.3)$ and $(\psi=0.01, \gamma=8, P_{_{R}}+P_{_{L}}=77)$. In the
right panel of Fig. \ref{ns-r-gf}, we have the allowed region in terms of $\epsilon$ and $\gamma$ which indicates that
\be\label{bound-epsilon-gamma}
\epsilon=(10^{-4} - 2\times 10^{-2}),\quad \gamma=(0.1 - 8).
\ee
Having the above results and after using the COBE normalization, we have
\be
\left(\frac{H}{\mpl}\right)^2=\frac{r~\pi^2}{P_{_{R}}+P_{_{L}}} \Delta_s^2=2\times 10^{-7}\epsilon,\nonumber
\ee
which determines the value of $H$ as
\be\label{H-value-gf}
H=(0.45 - 6.3)\times 10^{-5}\ \mpl\,.
\ee

Now, we can read $\kappa$ and $g$ in terms of the parameters $\gamma$ and $\epsilon$
\bse\label{psi-kappa-g}
\begin{align}
\gamma=\frac{g^2\psi^2}{H^2}\quad \Rightarrow\quad  \frac{g^2}{4\pi}=& 2\pi \mathcal{P}_{_{\mathcal{R}}}\gamma(\gamma+1)=1.5\times 10^{-8} \gamma(\gamma+1)\,,\\
\kappa g^2\psi^6 \simeq 2 \quad \Rightarrow \quad \kappa\simeq&\frac{2(1+\gamma)^2}{\epsilon^2H^2\gamma}=10^{7} \times \frac{(\gamma+1)^2}{\gamma\epsilon^3}.
\end{align}
\ese
From the left panel of Figs \ref{ns-r-gf}, we learn that in the allowed region the value of $\gamma$ is restricted as \eqref{bound-epsilon-gamma}, which determines the value of $\kappa$ and $g$
\be
g\simeq(0.15 - 3.7)\times 10^{-3},\qquad \Lambda\sim (10^{-5}-10^{-4})\mpl\,,\quad \kappa\equiv \Lambda^{-4}\,.
\ee
As an interesting and notable feature of our model, the value of the gauge coupling $g$, is directly related to the value of the power spectrum of CMB curvature fluctuations $\mathcal{P}_{_{\mathcal{R}}}$.

Restricting ourselves to 1$\sigma$ contour in the left panel of Fig. \ref{ns-r-gf}, we obtain the following bounds on $r$, $n_s$ and $H$
\be\label{2sigma-values}
0.98 \leq n_s\leq 0.99,\qquad 0.05<r<0.15,\qquad H\simeq (3.4-5.4)\times 10^{-5}\mpl,
\ee
which leads to the following bounds for $\psi$ and $\epsilon$ 
$$0.04\leq\psi\leq0.1\quad \textmd{and} \quad 0.6\times 10^{-2}\leq\epsilon\leq1.5\times 10^{-2},$$
where in 1$\sigma$ contour $0.5<\gamma<4$.
In our model, the field value $\psi$ is sub-Planckian and of order $10^{17}$ GeV, while the tensor to scalar ratio is considerable $r>0.02$. Thus, gauge-flation model satisfies a modified version of the Lyth bound \cite{Lyth-bound}.

\section{Summary, Discussion and Outlook}\label{Discussion-section}

In this work we have presented a detailed analysis of the gauge-flation model which we introduced in \cite{gauge-flation}. We first showed that non-Abelian gauge field theory can provide the setting for constructing an isotropic and homogeneous inflationary background. We did so by using the global part of the gauge symmetry of the problem and identified the $SU(2)$ subgroup of that with the rotation group. We argued that this can be done for \emph{any} non-Abelian gauge group, as any such group has an $SU(2)$ subgroup. Therefore, our discussions can open a new venue for building inflationary models, closer to particle physics high energy models, where non-Abelian gauge theories have a ubiquitous appearance.

The Yang-Mills theory cannot serve the job of building inflationary models, and we have to consider more complicated gauge theory actions. Among the obvious choices, we have checked non-Abelian version of Born-Infeld action \footnote{For an analysis of non-Abelian Born-Infled theory within the FRW cosmology see \cite{Galtsov}.} (with the symmetric trace prescription \cite{Tseytlin}), which does not lead to a slow-roll dynamics within its space of parameters. We have checked  $F^4$ terms which appear in one loop level effective gauge theory action. If we parameterize such $F^4$ terms as $\Tr(\alpha F^4+\beta  (F^2)^2)$, our analysis shows that it is  possible to get   slow-roll inflationary background for specific range of $\alpha$ and $\beta$ parameters. With the gauge group $SU(2)$, upon which  we have mainly focused in this work, the $\Tr(F\wedge F)^2$ that we have considered here can be obtained
from specific choices of $\alpha$ and $\beta$.

As discussed our motivation for considering a $\Tr(F\wedge F)^2$ term was primarily providing an explicit, simple realization of  our gauge-flation scenario which can lead to a satisfactory slow-roll inflation;  in this work we were not concerned with explicit derivation or embedding of this term from particle physics models. At technical level this happens because the dependence of this term on the background metric $g_{\mu\nu}$ appears only through $\det g$ and as a result the contribution of this term to the energy-momentum of the backfground will take the form of a perfect fluid with $P=-\rho$ equation of state, perfectly suited for driving an almost de Sitter expansion.
It is, however, important to study appearance of this $\kappa$-term through a rigorous quantum gauge field theory analysis and in particle physics settings. {}From  particle physics model building viewpoint, a $Tr(F\wedge F)^2$ type term
can  be argued for, considering  axions in a non-Abelian gauge theory \cite{Weinberg-vol2} and recalling the axion-gauge field interaction term ${\cal L}_{axion}\sim\frac{\varphi}{\Lambda} \Tr F\wedge F$. Then, integrating out the massive axion field $\varphi$ leads to an action of the form we have considered.
If we adopt this point of view our $\kappa$ parameter is then related to the cutoff scale $\Lambda$ as $\frac{\kappa}{384}\sim\Lambda^{-4}$ \cite{Weinberg-vol2}, and hence leading to $\Lambda\sim 10^{-4}\mpl\sim 10^{14}$ GeV. In order for this proposal to work, some points should be checked: recalling that $H\lesssim 10^{13}$ GeV, $\Lambda\sim 10 H$. For this one loop effective action description to make sense it is crucial that the cutoff $\Lambda$ becomes larger than $H$, because only axion configurations with subhorizon momenta $(k\gtrsim H)$ will contribute to
(quantum) loop corrections. The superhorizon modes, as in any quantum field theory on (almost) de Sitter background, are frozen and have become classical, and hence do not contribute to quantum corrections. It is also crucial that we are in a perturbative regime of the gauge theory with $g\sim 10^{-3}$. Therefore, we need not worry about complications of dealing with a confining (non-Abelian) gauge theory. In our case, we are in a weakly coupled regime where the theory is in deconfined phase.
We also remark that, as argued, during slow-roll inflation regime the contribution of the $\kappa$-term to the energy density of the gauge field configuration should dominate over that of the Yang-Mills part. In order for the mechanism for generation of the $\kappa$-term  sketched above to work, one should argue how the other possible higher-order terms, at $F^4$ level and higher loops (leading to higher powers of $F$ in the effective action), are suppressed compared to the $\kappa$-term. These issues will be discussed in a later publication.

Another interesting feature of our gauge-flation model is its naturalness; that demanding to have a successful inflationary model compatible with the current data leads to parameters which are within their natural range: the Hubble during inflation $H$ is of order $10^{1}$ GeV, and cutoff scale of the theory $\Lambda\sim 10^{14}-10^{15}$ GeV which are natural within the (SUSY) GUT models. Moreover, as is required by the consistency of the theory $H$ is less than cutoff $\Lambda$ (by one order of magnitude).
The other parameter of the theory, the gauge coupling $g\sim 10^{-3}-10^{-4}$, although a bit lower than the value expected for the coupling at the gauge unification scale, is also in a natural range. The field value $\psi_i$ and its displacement during inflation $\psi_i-\psi_f$, are both of order $10^{-2}\mpl$, well within the sub-Planckian regime. Therefore, as discussed, the arguments of standard single field inflationary models and the Lyth bound \cite{Lyth-bound} do not apply to our model and we do not face the super-Planckian field problem, which is a generic feature of large-field inflation models, such as chaotic inflation, causing concerns about the validity of using  classical Einstein gravity. We also note that the energy density during inflation $3H^2\mpl^2\sim  (2\times 10^{15}\ \mathrm{GeV})^4$, is the same order as the SUSY GUT scale.

Our other motivation for studying the gauge-flation scenario, which is at least  in spirit  close to beyond standard particle physics model settings,
was to provide a setup to address cosmological questions after inflation. As we discussed and is also seen from the phase diagram in Fig.~\ref{x=6.35-slow-roll-figures},  after the slow-roll ends  we enter a phase where the dynamics of the effective inflaton field, and gauge fields in general, is governed by the Yang-Mills term. The effective inflaton $\psi$ starts an oscillatory phase and through standard (p)reheating arguments, \eg see \cite{preheating-Linde}, it can lose its energy to the gauge fields. If we have an embedding of our gauge-flation scenario into beyond standard models, the energy of these gauge fields will then naturally be transferred to all the other standard model particles via standard gauge interactions. Therefore, our gauge-flation provides a natural setting for building (p)reheating models, to which we hope to return in future works.

\section*{Acknowledgements}

It is a pleasure to thank Niayesh Afshordi for
discussions and Amjad Ashoorioon, Robert Brandenberger, Masud Chaichian, Hassan Firouzjahi and Anca Tureanu  for comments on the draft. Work of A.M. is partially
supported by the grant from Bonyad-e Melli Nokhbegan of Iran.
\newpage

\appendix
\section{second-order action}
After a tedious but straightforward
calculation, which is also confirmed by the \texttt{Maple} codes,
the total action to second-order in perturbations is
\bea\label{scalar-2nd-action}
&~&~~~~~~\delta_{_{2}}S_{_{tot}}=\int ~ a^3 d^4x~ \bigg[\frac32\left(1+\kk\right)\frac{\dot{\mQ}^2}{a^2}-\bigg((1+\kk)\frac{k^2}{a^2}\frac{\tM}{\dot{\phi}/a}+3(1+\kk)\frac{\phi}{a}\dot\Psi\nonumber\\
&+&(1+\kk)\frac{\dot\phi}{a}(6\Psi+3\Phi)+4\frac{\kappa \dot\phi^2}{a^2}\frac{k^2}{a^2}\frac{\mM}{\dot{\phi}/a}\bigg)\frac{\dot \mQ}{a}+\bigg(3\frac{g^2\phi^2}{a^2}+12\frac{\dot\phi^2}{\phi^2}+3\dk\frac{g^2\phi^2}{a^2}-\frac{k^2}{a^2}\bigg)\frac{\mQ^2}{a^2}\nonumber\\
&+&\bigg(-12\kk\frac{\dot\phi}{a}\dot\Psi+2(3-2\dk)\frac{k^2}{a^2}\frac{\mM}{\phi/a}+(12\frac{g^2\phi^2}{a^2}(1-\dk)+2\frac{k^2}{a^2})\frac{\phi}{a}\Psi-6\frac{g^2\phi^2}{a^2}(1+\dk)\frac{\phi}{a}\Phi\nonumber\\&-&2\frac{\kappa g^2\phi^4}{a^4}
\frac{k^2}{a^2}\frac{\tM}{\phi/a}-2\frac{\kappa\dot\phi^2}{a^2}\frac{k^2}{a^2}\frac{\dot{\mM}}{\dot{\phi}/a}\bigg)\frac{\mQ}{a}-\frac32\frac{g^2\phi^4}{a^4H^2}\dot{\Psi}^2-3\frac{g^2\phi^4}{a^4H}\dot\Psi\Phi-\frac32\frac{g^2\phi^4}{a^4}\Psi^2-\frac32\frac{g^2\phi^4}{a^4}\Phi^2\nonumber\\
&+&\frac{k^2}{a^2}\Psi^2-\frac{2k^2}{a^2}\Psi\Phi+\frac{k^2}{a^2}\frac{\dot{\mM}^2}{g^2\phi^4/a^4}+2\dk\frac{k^2}{a^2}\frac{\phi}{\dot{\phi}}\Psi\dot{\mM}+\left(H^2(2-\epsilon)+\frac{k^2}{a^2}(-1+\frac23\dk)\right)\frac{k^2}{a^2}\frac{\mM^2}{g^2\phi^4/a^4}
\nonumber\\&+&\left(\frac{g^2\phi^2}{\dot{\phi}^2}+\frac{1}{2\dot{\phi}^2/a^2}(1+\frac13\kk)\frac{k^2}{a^2}\right)\frac{k^2}{a^2}\tM^2+\bigg((1+\kk)(\frac{\phi}{\dot\phi}\dot{\Psi}+\Phi)+(2-\epsilon\YM2)\Psi-2\frac{\dot \mM}{\phi\dot\phi}\nonumber\\
&+&2\frac{\mM}{\phi^2/a^2}+\frac23\dk\frac{k^2}{a^2}\frac{\mM}{\dot\phi^2/a^2}\bigg)\frac{k^2}{a^2}\tM+
\left(4\dk\frac{\phi}{\dot\phi}\dot\Psi+4(-1+\dk)\Psi+2(1+\dk)\Phi\right)\frac{k^2}{a^2}\mM\bigg].\nonumber
\eea
Note that the above action is computed  fixing the Newtonian gauge $E=B=0$.
The field equations of $\mM$ and $\tM$ reduce to \eqref{A0eq}. On the other hand, using  the constraint \eqref{A0eq}, the field equation of $\mQ$ is equal to the energy conservation equation \eqref{energy-conservation}. The field equation of $\Psi$ is identical to \eqref{dP} plus the field equation of $\mQ$, while the field equation of $\Phi$ is equal to \eqref{drho}.

\section{Short review of constraint structure in cosmic perturbation theory}
Consider the \emph{scalar} sector of metric perturbations \eqref{metric-pert}, parameterized as
\be\label{m}%
\begin{split}
ds^2=-(1+2A)dt^2+a^2\left((1-2C)\delta_{ij}+2\partial_{ij}E\right)dx^idx^j\,,
\end{split}
\ee
where $\partial_i$ denotes partial derivative respect to $x^i$.
For multi-scalar field inflationary models, minimally coupled to gravity, generally we have
\bea\label{a1}
\nonumber\\
\delta\rho&=&\sum_I\big(\dot\varphi_I(\delta\dot\varphi_I-A\dot\varphi_I)+V_{,I}\delta\varphi_I\big),\nonumber\\
\delta P&=&\sum_I\big(\dot\varphi_I(\delta\dot\varphi_I-A\dot\varphi_I)-V_{,I}\delta\varphi_I\big),\\
\delta q&=&-\sum_I\dot{\varphi}_I\delta\varphi_I,\nonumber\\
\pi^s&=&0,\nonumber
\eea
where $V_{,I}$ is equal to $\frac{\partial V}{\partial\varphi_I}$ and for simplicity we chose the Newtonian gauge $(E=B=0)$.
Also, the gravitational field equations are as below, two constraints
\bea\label{a2}
&~&\delta q+2(\dot C+HC)=0,\\
\label{a3}
&~&\delta\rho-3H\delta q+2\frac{k^2}{a^2}C=0,\\
\label{a-pi}
&~&a^2\pi^S=C-A,
\eea
and a dynamical equation
\bea\label{a4}
\delta P+\delta\dot q+3H\delta q+(\rho_0+P_0)C=0.
\eea
Since in all the scalar driven models $a^2\pi^S$ is identically zero, for this systems we have
\be
C=A.
\ee

From the combination of \eqref{a1}, \eqref{a2} and \eqref{a4}, we obtain
\bea
\sum_I\big(\ddot\varphi_I+3H\dot\varphi_I+V_I\big)=0,
\eea
which is a summation of the background field equations
\bea
\forall \varphi_I:~~~~\ddot\varphi_I+3H\dot\varphi_I+V_I=0,
\eea
That is, \eqref{a2} is not an independent equation and only the constraint \eqref{a3} and the dynamical equation \eqref{a4} are independent among the Einstein equations.
At last in these models, the complete set of the equations  consists of constraint \eqref{a3}, and $n$ field equations of $\delta\varphi_I$. This makes it possible to decompose each arbitrary field perturbation into
an adiabatic (curvature) perturbation along the inflaton trajectory and
$n-1$ entropy (isocurvature) perturbations orthogonal to the inflaton in
field space \cite{two-field-models}.

In contrast to the multi-scalar field models, gauge-flation has three independent gravitational field equations: \emph{three constraints} \eqref{a2}-\eqref{a-pi}, and a dynamical equation \eqref{a4} as well as the gauge field constraint  ($D_\mu\frac{\partial\mathcal{L}}{\partial F^a_{\mu0}}=0$) for five unknowns.
In fact, since in gauge-flation we have a non-Abelian gauge field and not simply some scalar fields, the standard approach for dealing with multi-scalar fields is not applicable even in the study of scalar sector of gauge-flation perturbation theory.


\end{document}